\documentclass{aa}  
\usepackage{natbib}

\usepackage{graphicx}
\usepackage{txfonts}
\begin{document}

   \title{Stellar systems in the direction of Pegasus\,I - I. Low surface 
brightness galaxies}

   \author{N\'elida M. Gonz\'alez
          \inst{1,2},
	  Anal\'ia V. Smith Castelli\inst{1,2,3},
          Favio R. Faifer\inst{1,2,3},
         Carlos G. Escudero\inst{1,2}
\and
         Sergio A. Cellone\inst{1,3,4}
          }

   \institute{Facultad  de  Ciencias  Astron\'omicas  y  Geof\'isicas,
     Universidad Nacional de La Plata, Paseo del Bosque , B1900FWA,
     La Plata, Argentina
         \and
             Instituto  de Astrof\'isica  de La  Plata (UNLP-CONICET),
             Paseo del Bosque, B1900FWA, La Plata, Argentina\\
             \email{ngonzalez@fcaglp.unlp.edu.ar}
        \and
	      Consejo  Nacional  de  Investigaciones  Cient\'ificas  y
              T\'ecnicas, Godoy Cruz 2290, C1425FQB, CABA, Argentina
        \and     
        Complejo Astronómico  El Leoncito  (CONICET - UNLP - UNC - UNSJ),
        Av. Espa\~na 1512 (Sur), J5402DSP, San Juan, Argentina
}

   \date{Received September 1, 2018; accepted September 27, 2018}

 \abstract
   {In spite  of the  numerous studies  of low-luminosity  galaxies in
     different environments,  there is still no  consensus about their
     formation  scenario. In  particular, a  large number  of galaxies
     displaying extremely low-surface  brightnesses have been detected
     in the last years, and the nature of these objects is still under
     discussion. }
   {To  enlarge  the  sample  of known  low-surface  brightness  (LSB)
     galaxies  and to  try to  provide  clues about  their nature,  we
     report the detection of eight  of this type of objects ($\mu_{\mathrm{
       eff}, g'} \simeq 27$  mag~arcsec$^{-2}$)  towards  the  group  of
     galaxies Pegasus\,I.  They are  located, in projection,  within a
     radius of $\sim200$\,kpc in the very center  of Pegasus\,I, close
     to the dominant elliptical galaxies NGC\,7619 and NGC\,7626.}
   {We analyzed  deep, high-quality,  GEMINI-GMOS images  with ELLIPSE
     within  IRAF in  order to  obtain their  brightness profiles  and
     structural parameters.   We also fit S\'ersic  functions to these
     profiles in order to compare  their properties with those of {\it
       typical} early-type galaxies.} 
   {Assuming that these galaxies are at the distance of Pegasus\,I, we
     have  found  that  their  sizes are  intermediate  among  similar
     objects reported in the literature.  In particular, we found that
     three  of  these  galaxies  can be  classified  as  ultra-diffuse
     galaxies and a  fourth one displays a nucleus. The  eight new LSB
     galaxies  show  a  significant color  dispersion around  the
     extrapolation towards  faint luminosities of  the color-magnitude
     relation  defined  by  {\it  typical}  early-type  galaxies.   In
     addition,  they display  values  of the  S\'ersic  index below  1
     (concave brightness profiles in  linear scale), in agreement with
     values obtained for LSB galaxies in other environments.}
   {We also  show that  there seems to  be a bias  effect in  the size
     distributions of  the detected LSBs in  different environments, in
     the   sense  that   more  distant   groups/clusters  lack   small
     $r_\mathrm{eff}$ objects,  while large  systems are not  found in
     the Local Group  and nearby environments.  While there  may be an
     actual shortage of large LSB galaxies in low-density environments
     like  the Local  Group, the  non-detection of  small (and  faint)
     systems at large  distances is clearly a selection  effect. As an
     example,  LSB  galaxies  with  similar  sizes  to  those  of  the
     satellites of  Andromeda in  the Local  Group, will  be certainly
     missed in a visual identification at the distance of Pegasus\,I.}

   \keywords{methods:  observational  --  techniques:  photometric  --
     galaxies:  groups:  individual:   Pegasus\,I  --  galaxies:  star
     clusters: general  -- galaxies: dwarfs 
               }

\authorrunning{Gonz\'alez et al.}
\titlerunning {Stellar systems in the direction of Pegasus\,I - I. LSB
  galaxies}
   \maketitle
%

\section{Introduction}
\label{introduccion}

The  {\it   ``cloud  of  nebulae''}  extending   through  the  Pegasus
constellation
\citep[Pegasus\,I;][]{1942PASP...54..185Z,1965cgcg.book.....Z}     was
initially  reported to  be a  {\it  ``medium compact  cluster with  an
  angular diameter  of 6.3  degrees''}, containing  $\sim30$ confirmed
members,  with a  similar number  of spiral  and early-type  galaxies,
spanning  a   radial  velocity   range  of   $2500-5500$  km\,s$^{-1}$
\citep[][and                                                references
  therein]{1976PASP...88..388C,1980ApJ...239...12A}.   From   an  H\,I
study    of     80    galaxies     in    the     Pegasus\,I    region,
\citet{1982A&A...109..155R} increased the  number of confirmed members
to  75  bright galaxies  located  at  51.8\,Mpc,  with a  mean  radial
velocity  of 3885\,km\,s$^{-1}$  and  a small  velocity dispersion  of
236\,km\,s$^{-1}$. This  distance is  in agreement  with the  value of
53\,Mpc obtained more  recently by \citet{2001ApJ...546..681T} through
the application of the surface brightness fluctuations method.

The Pegasus\,I group is dominated  by two giant elliptical galaxies of
similar   luminosities:  NGC\,7619   ($V=11.06$  mag)   and  NGC\,7626
($V=11.08$\,mag) \citep{1992yCat.7137....0D}. Both  galaxies are radio
sources  and  show  morphological peculiarities.   NGC\,7626  presents
symmetric radio-jets/lobes on each side of its central core, outer and
inner  shells, a  nuclear dust  lane, a  compact X-ray  source at  its
center,               among               other               features
\citep{1985ApJ...291...32B,1988ApJ...330L..87J,1993A&A...279..376B,1997A&A...318..361T}.
NGC\,7619, in turn, is a strong  and extended X-ray source with a long
asymmetric     X-ray    tail     in     the    southwest     direction
\citep{2008ApJ...688..931K,2009ApJ...696.1431R}.   This tail  has been
interpreted as  due to the infall  of this galaxy into  Pegasus\,I. In
addition,  cold fronts  detected in  each of  these dominant  galaxies
using  {\it Chandra}  images  \citep{2009ApJ...696.1431R} support  the
scenario of a major merger of two subgroups. It is worth noticing that
Pegasus\,I   was  earlier   reported  as   presenting  an   excess  of
star-forming early-type galaxies in its outskirts, like Butcher-Oemler
clusters \citep{1989AJ.....98.2044V,1997MNRAS.286..133T}.  This excess
would be associated with the infall of galaxies into Pegasus\,I, which
would  indicate   that  it   is  a   group  in   formation  \citep[see
  also,][]{2003PASJ...55..593F}.

  \citet{1986ApJ...304..312C} report  an intergalactic gas  density of
  $2\times10^{-4}$\,cm$^{-3}$ for  Pegasus\,I, i.e., one of  the least
  dense media  ever detected in a  group of galaxies.  From  this fact
  and the low-velocity  dispersion of the group, it  would be expected
  that ram-pressure stripping does not play  a key role in driving its
  galaxy evolution.   However, \citet{2007AJ....133.1104L}  have found
  that, though  as not as  intense as in  rich clusters, there  is gas
  depletion in a significant number  of late-type galaxies. This would
  indicate that  the interaction  between the interstellar  medium and
  the intracluster medium  may be indeed playing an  important role in
  Pegasus\,I.

  The  faint galaxy  content of  Pegasus\,I has  been poorly  studied.
  \citet{1994AJ....108.1209V}  presented surface  photometry of  seven
  galaxies displaying dwarf morphologies  located in the neighborhoods
  of  NGC\,7626,  but with  no  radial  velocity information  at  that
  moment.  A few years  later, \citet{1997MNRAS.286..133T} attempted a
  determination of the faint end  of the galaxy luminosity function in
  Pegasus\,I; however,  the low galaxy  density in his  sampled fields
  resulted in a relatively small  number of objects brighter than $M_R
  > -11$,  thus  leaving  the group  luminosity  function  essentially
  unconstrained.
More recently, through a friends--of--friends (FoF) algorithm,
\citet{2002AJ....123.2976R} identified Pegasus\,I as a physical association
with 13 spectroscopically confirmed members.

In this context, we have started the study of the faint galaxy content
of  the  central region  of  Pegasus\,I  through deep  optical  images
obtained with the  telescope of the Gemini North  Observatory. In this
paper (the first of a series)  we present a photometric study of eight
low-surface-brightness  (LSB) galaxies  detected  towards the  central
region of Pegasus\,I.  Originally identified  in the Local Group (LG),
LSB  galaxies   are  extended   objects  displaying   central  surface
brightnesses             $\mu_{0,B}>23$             mag\,arcsec$^{-2}$
\citep{1997ARA&A..35..267I}. This characteristic makes their detection
and  subsequent analysis  a  challenging task.   The  interest in  the
detection  of  such  extremely  faint  galaxies  outside  the  LG  has
increased in  the last years,  as they  are expected to  impose strong
constraints  to  models  of  galaxy  formation  and  evolution.   Some
questions arise when analyzing  these objects: are these extragalactic
systems the  counterparts of the  LG dwarf spheroidal  (dSph) galaxies
which are supposed  to be the most dark matter  (DM) dominated systems
in  the  near  Universe?   Are   they  formed  in-situ,  or  are  they
kinematically decoupled  and gravitationally bound  structures arising
from the interaction  of massive galaxies that, as  a consequence, are
not   expected  to   contain   DM  at   all  \citep[][and   references
  therein]{2013MNRAS.429.1858D}?   Previous efforts  in this  sense in
Pegasus\,I    have   been    made,   with    small   telescopes,    by
\citet{1997AJ....114.2448O} and  \citet{2017ApJ...846...26S}.  None of
the  galaxies presented  in  this  paper has  been  included in  those
works. Therefore, our sample contributes to increase the number of LSB
galaxies identified towards Pegasus\,I.

The paper is organized  as follows. In Section\,\ref{observaciones} we
describe  the  photometric  data.   In  Sections\,\ref{photometry}  and
\ref{profiles},   the  strategies   followed  to   obtain  photometric
parameters. In Section\,\ref{Samples} we  present the results obtained
from our  analysis and, in Section\,\ref{discusion},  a discussion and
our conclusions.
\section{Observational data}
\label{observaciones}

\begin{figure}[th!]
\centering
\includegraphics[scale=0.53]{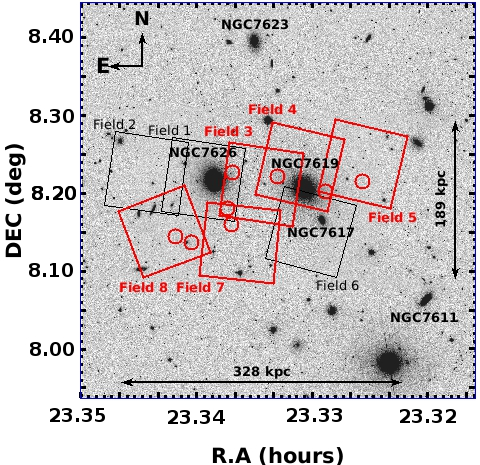}
\caption{30.5 $\times$ 30.5 arcmin mosaic  in the $r'$ filter obtained
  from  the  SDSS  database  (DR12)  showing  the  central  region  of
  Pegasus\,I.  Red squares depict the  Gemini-GMOS frames used in this
  work ($\sim 5.5$  arcmin on a side).   Their designation corresponds
  to the  order in  which each  frame was  observed during  our Gemini
  survey of Pegasus\,I.  Red circles  indicate the location of the LSB
  objects. The large  arrows depict the spatial scales  covered by our
  Gemini-GMOS fields. North is up and East to the left.}
\label{SDSS}  
\end{figure}

We have obtained, deep $g'$, $r'$,  and $i'$ images of eight fields in
the  central  region  of  Pegasus  I,  with  the  Gemini  Multi-Object
Spectrograph (GMOS; Hook et al. 2004) of GEMINI-North, along semesters
2008B, 2012A, 2014A, 2014B and  2015B.  These observations are part of
an ongoing survey  of the globular clusters population  in the central
region of  Pegasus\,I with  the Gemini telescopes  (Faifer et  al., in
preparation).   In  five  of  those fields  we  identified  eight  LSB
galaxies.  Figure\,\ref{SDSS} shows the location of the eight observed
fields, and the projected spatial distribution of the detected objects
in the central region of Pegasus\,I.

GMOS consists of 3 CCDs with 2048 $\times$ 4096 pixels each, separated
by $\sim$ 2.8 arcsec gaps, and  with an unbinned pixel scale of 0.0727
arcsec  pixel$^{-1}$.  The  field of  view (FOV)  is 5.5  $\times$ 5.5
arcmin  and  the scale  for  binning  $2  \times  2$ is  0.146  arcsec
pixel$^{-1}$.  In Table\,\ref{info} we show the information related to
the GEMINI-GMOS images.
\begin{table*}[th!]
\caption{Log of the GEMINI-GMOS images.}
\label{info}
\centering
\begin{tabular}{cccccccc}
\hline\hline
\multicolumn{1}{c}{Field}&\multicolumn{1}{c}{Program}&\multicolumn{1}{c}{R.A}&\multicolumn{1}{c}{DEC}&\multicolumn{1}{c}{Filter} & \multicolumn{1}{c}{Date} & \multicolumn{1}{c}{Exposures} & \multicolumn{1}{c}{FWHM}\\
\multicolumn{1}{c}{} & \multicolumn{1}{c}{} &  (J2000) &(J2000)&\multicolumn{1}{c}{} & \multicolumn{1}{c}{} & \multicolumn{1}{c}{(sec)} & \multicolumn{1}{c}{(arcsec)}  \\
\hline
  & 		  &             &	     & $g'$ &  4 Aug. 2008 & 8$\times$450 & 0.9\\
1 & GN-2008B-Q-14 & 23:20:46.13 & 08:13:05.9 & $r'$ &  4 Aug. 2008 & 4$\times$450 & 0.8\\
  & 		  &             &	     & $i'$ &  4 Aug. 2008 & 4$\times$450 & 0.8\\
\hline
  & 		  &             &	     & $g'$ &  8 Aug. 2008 & 8$\times$450 & 0.8\\
2 & GN-2008B-Q-14 & 23:21:03.85 & 08:13:31.6 & $r'$ &  8 Aug. 2008 & 4$\times$450 & 0.7\\
  & 		  &             &	     & $i'$ &  8 Aug. 2008 & 4$\times$450 & 0.7\\
\hline
  & 		  &             &	     & $g'$ & 10 Aug. 2008 & 8$\times$450 & 0.8\\
3 & GN-2008B-Q-14 & 23:20:27.82 & 08:12:41.3 & $r'$ & 10 Aug. 2008 & 4$\times$450 & 0.8\\
  & 		  &             &	     & $i'$ & 10 Aug. 2008 & 4$\times$450 & 0.9\\
\hline
  & 		  &             &	     & $g'$ & 8 Aug. 2008 & 8$\times$450 & 0.9\\
4 & GN-2008B-Q-14 & 23:20:15.62 & 08:14:00.8 & $r'$ & 8 Aug. 2008 & 4$\times$450 & 1.0\\
  & 		  &             &	     & $i'$ & 8 Aug. 2008 & 4$\times$450 & 0.7\\
\hline
  & 		  &             &	     & $g'$ & 23 Jun. 2012 & 8$\times$440 & 0.8\\
5 & GN-2012A-Q-55 & 23:19:56.08 & 08:14:14.9 & $r'$ & 21 Jun. 2012 & 4$\times$440 & 0.7\\
  & 		  &             &            & $i'$ & 24 Jun. 2012 & 6$\times$300 & 0.8\\
\hline
  & 		  &             &	     & $g'$ & 23 Jun. 2014 & 8$\times$450 & 0.7\\
6 & GN-2014A-Q-70 & 23:20:12.68 & 08:08:58.91& $r'$ & 25 Jun. 2014 & 4$\times$450 & 0.6\\
  & 		  &             &            & $i'$ & 25 Jun. 2014 & 6$\times$300 & 0.5\\
\hline
  & 		  &             &	     & $g'$ & 27 Jul. 2014 & 8$\times$450 & 1.0\\
7 & GN-2014B-Q-17 & 23:20:34.10 & 08:08:06.1 & $r'$ & 27 Jul. 2014 & 4$\times$450 & 0.9\\
  & 		  &             &            & $i'$ & 27 Jul. 2014 & 6$\times$300 & 0.9\\
\hline
  & 		  &             &            & $g'$ & 25 Jul. 2015 & 10$\times$450 & 0.6\\
8 & GN-2015B-Q-13 & 23:20:58.08 & 08:09:03.9 & $r'$ & 25 Jul. 2015 &  8$\times$300 & 0.6\\
  & 		  &             &            & $i'$ & 25 Jul. 2015 &  8$\times$300 & 0.5\\
\hline
\end{tabular}
\end{table*}
\begin{table*}[th!]
\caption{Distance information of the dominant galaxies of Pegasus\,I.}
\label{distancias}
\centering
\begin{tabular}{lccccccc}
\hline\hline
\multicolumn{1}{c}{Galaxy} & \multicolumn{1}{c}{R.A.} & \multicolumn{1}{c}{DEC} & \multicolumn{1}{c}{($m-M$)} & \multicolumn{1}{c}{Distance} & \multicolumn{1}{c}{Method} & \multicolumn{1}{c}{$\rm H_0$} & \multicolumn{1}{c}{Reference} \\
\multicolumn{1}{c}{} & \multicolumn{1}{c}{(J2000)} &\multicolumn{1}{c}{(J2000)} & \multicolumn{1}{c}{(mag)} & \multicolumn{1}{c}{(Mpc)} & \multicolumn{1}{c}{} & \multicolumn{1}{c}{(km s$^{-1}$ Mpc$^{-1}$)}  & \multicolumn{1}{c}{}  \\
\hline
NGC\,7619       & 23:20:14.5 & 08:12:22 & 33.70  & 55.0 & SBF   & 74.4 & \citet{2013AJ....146...86T}  \\
NGC\,7626       & 23:20:42.5 & 08:13:01 & 33.34  & 46.6 & SBF   & 74.4 & \citet{2013AJ....146...86T}  \\
\hline
\end{tabular}
\tablefoot{SBF: Surface brightness fluctuations.}
\end{table*}

We performed spatial dithering between the individual exposures 
 in order  to facilitate the removal of cosmic  rays and to fill
in the  gaps between  the CCD  chips.  The  raw images  were processed
using  the   GMOS  package   within  IRAF   (e.g.   \textsf{gprepare},
\textsf{gbias},      \textsf{giflat},       \textsf{gireduce}      and
\textsf{gmosaic}).  The  appropriate bias  and flat-field  images were
obtained from the Gemini Science Archive (GSA) as part of the standard
GMOS baseline calibrations.  

As  previously mentioned,  the data  used in  this work  were obtained
along several  semesters. For that  reason, some fields  were observed
with  the  GMOS-N  E2V  DD  CCDs (previous  to  their  replacement  by
Hamamatsu CCDs), thus displaying strong  fringing patterns in the $i'$
frames.  These  patterns were removed  by subtracting a  master fringe
frame that was created using \textsf{gifringe} and \textsf{girmfringe}
task.

The resulting  images corresponding  to the same  filter and  the same
field obtained from  the above procedure, were  finally co-added using
the  task \textsf{imcoadd}  in order  to  obtain the  images used  for
further analysis.  The photometric data  obtained from them were later
calibrated to  the photometric  system of the  {\it Sloan  Digital Sky
  Survey} (SDSS).
\section{Sky subtraction and detection of LSB galaxies}
\label{photometry}
From visual inspection  of our GEMINI-GMOS images  we identified eight
diffuse and  extended objects (Table\,\ref{COG}).  In  order to obtain
their brightness profiles and their  photometric parameters, we had to
subtract the contribution  of the sky to the images.   It is important
to note that  the average surface brightness of the  sky in our images
is      $\mu_{\mathrm{sky},g'}      \simeq      22$~mag~arcsec$^{-2}$,
$\mu_{\mathrm{sky},r'}          \simeq          21$~mag~arcsec$^{-2}$,
$\mu_{\mathrm{sky},{i'}}   \simeq  20$~mag~arcsec$^{-2}$,   while  the
objects of our study display central surface brightnesses much fainter
than   these   values   ($\mu_{g'}   \gtrsim   25~$mag~arcsec$^{-2}$).
Therefore, to  get reliable photometric and  structural parameters, it
is  necessary to  perform a  careful modeling  and subtraction  of the
background.   The general  process consisted  of modeling  the surface
brightness  distributions of  the  elliptical  galaxies NGC\,7626  and
NGC\,7619  and  their  respective halos,  including  several  extended
objects  that  could  affect  the surface  brightness  profiles.   One
difficulty with this process was that  the halos of the two elliptical
galaxies  extended beyond  the  edge of  each  individual field.   The
different  strategies followed  in  each field  to  achieve this  goal
included:

\begin{itemize}
\item[]  {\bf Field  3 -}  {\it The  subtraction of  the halos  of the
  bright elliptical  galaxies NGC\,7619  and NGC\,7626,  combined with
  SExtractor.}  After  subtracting preliminary constant values  of the
  sky level  from the  original images,  we modeled  the halos  of the
  galaxies     through     the     task    ELLIPSE     within     IRAF
  \citep{1986SPIE..627..733T,1987MNRAS.226..747J}.  These  models were
  subtracted from  the sky-subtracted images, but  the residuals still
  showed significant systematic  variations in the regions  of the LSB
  galaxies.  As  the photometric parameters  of this type  of galaxies
  are   very  sensitive   to  sky   variations,  we   used  SExtractor
  \citep{1996A&AS..117..393B}  to  build  a  model  of  this  residual
  background,   which    was   then   subtracted    from   the
    halos-subtracted frames.
 
\item[] {\bf Field 4 -} {\it The subtraction of models of NGC\,7619,
    extended objects and bright stars present in the field.}  After
  subtracting a preliminary mean sky value from the images, we
  modeled the light of NGC\,7619 through ELLIPSE.  Other extended
    objects and bright stars, in the region of the detected LSB galaxies,
    were also modeled and subtracted in order to diminish any remaining
  background variation.  After a couple of iterations in the
    modeling-subtracting process, we obtained satisfactorily flat frames (the mean
  residual value is lower than 0.5 \% of the
  original value) in the regions of the LSB galaxies.\\

\item[] {\bf Field 5 -} {\it  The subtraction of the halo of NGC\,7619
  combined with SExtractor.} The procedure in this case was the same
  as for field 3. \\

\item[]  {\bf  Field  7  -}  {\it  The  subtraction  of  the  halo  of
  NGC\,7626.}  We  used here  a similar procedure  to that  applied to
  field 3 but, in this case,  SExtractor was not needed to improve the
  background behavior  in the  regions of the  LSB galaxies  (the mean
  residual value is lower than 0.25 \% of the original value)\\

\item[] {\bf Field 8 -} {\it The subtraction of the models of extended
  objects located near the LSB  galaxies.}  We worked similarly as for
  field 4, excepting that these images  were not affected by the halos
  of the bright ellipticals.\\
\end{itemize}

\section{Surface brightness profiles}
\label{profiles}
\subsection{Model-independent parameters}
\label{independent}
\begin{table*}[h!]
\caption{Model-independent photometric  parameters of  the low-surface
  brightness objects presented in this  paper. They were obtained from
  the analysis of the curves of growth (COG).  Apparent magnitudes and
  colors are not extinction nor reddening corrected.}
\centering
\begin{tabular}{ccccccccc}
\hline 
\hline
Object & R.A.    & DEC    & Filter & $r_{\rm tot}$ & $r_{\rm eff}$ & $m_{\rm tot}$  & ($g'$-$r'$) & $\langle{\mu}_{{\rm eff}}\rangle$\\
       & (J2000) &(J2000) &        &  (arcsec)   &    (arcsec) & (mag)  & (mag)      & (mag~arcsec$^{-2}$)\\
  (1)  & (2)     & (3)    & (4)    &  (5)        &  (6)        & (7)    & (8)        & (9)\\
\hline
\hline\\
\medskip

                     &          &            & $g'$ &      & 2.52 $_{-0.28}^{+0.28}$ & 22.53 $_{-0.28}^{+0.20}$ &                      & 26.53 $_{-0.39}^{+0.30}$\\
\medskip
PEG J231956+081253.7 & 23:19:56 & 08:12:53.7 & $r'$ & 7.15 & 2.66 $_{-0.14}^{+0.14}$ & 21.69 $_{-0.15}^{+0.16}$ & 0.83 $_{-0.31}^{+0.26}$ & 25.82 $_{-0.19}^{+0.19}$
\\ 
\medskip
                     &          &            & $i'$ &      & 2.66 $_{-0.42}^{+0.42}$ & 21.48 $_{-0.29}^{+0.33}$ &                      & 25.60 $_{-0.50}^{+0.44}$ 
\\
\hline\\
\medskip

                     &          &            & $g'$ &       & 7.78 $_{-2.39}^{+4.79}$ & 21.09 $_{-0.65}^{+0.92}$ &                      & 27.54 $_{-1.22}^{+1.27}$    \\
\medskip
PEG J232023+081331.4 & 23:20:23 & 08:13:31.4 & $r'$ & 19.15 & 8.38 $_{-1.80}^{+2.39}$ & 19.97 $_{-0.47}^{+0.71}$ & 1.11 $_{-0.80}^{+1.17}$ & 26.58 $_{-0.77}^{+0.86}$ 
\\ 
\medskip
                     &          &            & $i'$ &       & 7.78 $_{-1.19}^{+2.99}$ & 19.70 $_{-0.50}^{+0.63}$ &                      & 26.16 $_{-0.65}^{+0.88}$ 
\\
\hline
{\tiny} &\\
\medskip
                     &          &            & $g'$ &       & 8.30 $_{-0.71}^{+1.18}$ & 20.27 $_{-0.34}^{+0.45}$ &                      & 26.86 $_{-0.39}^{+0.52}$ \\
\medskip
PEG J232024+081209.0 & 23:20:24 & 08:12:09.0 & $r'$ & 16.35 & 7.58 $_{-1.18}^{+2.13}$ & 19.53 $_{-0.44}^{+0.62}$ & 0.74 $_{-0.55}^{+0.77}$ & 25.93 $_{-0.60}^{+0.79}$ \\ 
\medskip
                     &          &            & $i'$ &       & 7.82 $_{-0.95}^{+1.89}$ & 19.66 $_{-0.44}^{+0.61}$ &                      & 26.12 $_{-0.53}^{+0.75}$ 
\\
\hline\\
\medskip
                     &          &            & $g'$ &       & 5.67 $_{-1.16}^{+1.54}$ & 20.34 $_{-0.35}^{+0.44}$ &                      & 26.10 $_{-0.67}^{+0.64}$ \\
\medskip
PEG J232037+080934.3 & 23:20:37 & 08:09:34.3 & $r'$ & 16.94 & 5.02 $_{-1.28}^{+1.20}$ & 19.93 $_{-0.35}^{+0.40}$ & 0.40 $_{-0.50}^{+0.60}$ & 25.43 $_{-0.85}^{+0.59}$  
\\ 
\medskip
                     &          &            & $i'$ &       & 5.02 $_{-1.28}^{+0.89}$ & 19.60 $_{-0.31}^{+0.35}$ &                      & 25.10 $_{-0.83}^{+0.48}$ 
\\
\hline\\
\medskip
                     &          &            & $g'$ &       & 2.17 $_{-0.33}^{+0.24}$ & 22.36 $_{-0.27}^{+0.21}$ &                      & 26.04 $_{-0.48}^{+0.30}$  \\
\medskip
PEG J232037+081336.6 & 23:20:37 & 08:13:36.6 & $r'$ & 7.17  & 2.35 $_{-0.15}^{+0.18}$ & 21.62 $_{-0.15}^{+0.12}$ & 0.74 $_{-0.30}^{+0.24}$ & 25.47 $_{-0.21}^{+0.20}$ \\ 
\medskip
                     &          &            & $i'$ &       & 2.50 $_{-0.52}^{+0.43}$ & 21.16 $_{-0.33}^{+0.37}$ &                      & 25.16 $_{-0.68}^{+0.49}$ 
  \\
\hline\\
\medskip
                     &          &            & $g'$ &       & 2.72 $_{-0.82}^{+0.77}$ & 21.53 $_{-0.33}^{+0.38}$ &                      & 25.70 $_{-1.05}^{+0.59}$  \\
\medskip
PEG J232038+081046.9 & 23:20:38 & 08:10:46.9 & $r'$ & 9.26  & 2.72 $_{-1.09}^{+1.06}$ & 20.99 $_{-0.40}^{+0.33}$ & 0.53 $_{-0.52}^{+0.47}$ & 25.16 $_{-1.79}^{+0.71}$ \\ 
\medskip
                     &          &            & $i'$ &       & 2.99 $_{-1.09}^{+1.33}$ & 20.62 $_{-0.45}^{+0.40}$ &                      & 25.00 $_{-1.48}^{+0.80}$ 
\\
\hline\\
\medskip
                     &          &            & $g'$ &       & 3.07 $_{-0.92}^{+0.92}$ & 22.02 $_{-0.33}^{+0.40}$ &                      & 26.45 $_{-1.05}^{+0.65}$\\
\medskip
PEG J232049+080806.2 & 23:20:49 & 08:08:06.2 & $r'$ & 9.71  & 2.96 $_{-1.43}^{+1.43}$ & 21.53 $_{-0.50}^{+0.61}$ & 0.49 $_{-0.60}^{+0.73}$ & 25.88 $_{-3.69}^{+0.96} $
\\ 
\medskip
                     &          &            & $i'$ &       & 3.78 $_{-1.02}^{+1.43}$ & 20.80 $_{-0.36}^{+0.48}$ &                      & 25.67 $_{-0.91}^{+0.78}$ 
\\
\hline\\
\medskip
                     &          &            & $g'$ &       & 3.99 $_{-1.06}^{+1.99}$ & 21.69 $_{-0.53}^{+0.58}$ &                      & 26.70 $_{-0.98}^{+0.95}$ \\
\medskip
PEG J232054+080838.8 & 23:20:54 & 08:08:38.8 & $r'$ & 10.93 & 3.99 $_{-0.80}^{+1.33}$ & 21.10 $_{-0.40}^{+0.46}$ & 0.59 $_{-0.66}^{+0.74}$ & 26.11 $_{-0.68}^{+0.72}$   \\ 
\medskip
                     &          &            & $i'$ &       & 4.13 $_{-1.06}^{+2.13}$ & 20.70 $_{-0.57}^{+0.67}$ &                      & 25.78 $_{-0.97}^{+1.02}$\\
\hline
\hline
\end{tabular}
\tablefoot{\\
{\it   Col   (5):}   $r_{\rm    tot}$   is   the   equivalent   radius
($r=\sqrt{ab}=a\sqrt{1-\epsilon}$)  at  which   the  curve  of  growth
stabilizes in the three filters.\\  {\it Col (6):} $r_\mathrm{eff}$ is
the  equivalent  radius   that  contains  half  of   the  flux  within
$r_\mathrm{tot}$.\\  {\it Col  (7):} Total  integrated magnitudes  are
calculated from  the numerical integration of  the observed brightness
profiles     up    to     $r_\mathrm{tot}$.\\    {\it     Col    (9):}
$\langle\mu_\mathrm{eff}\rangle       =       mag+2.5~\log(2       \pi
r_\mathrm{eff}^2)$.  }
\label{COG}
\end{table*}
 
In  order  to obtain  the  brightness  profiles  of the  LSB  galaxies
detected in  our frames,  we worked  with ELLIPSE  within IRAF  on the
GEMINI-GMOS sky-subtracted and registered  $g'$, $r'$ and $i'$ images.
In  all  the   images,  these  objects  display   smooth  and  diffuse
morphologies  without showing  inner substructure  (bars, dust  bands,
etc.) or color gradients in their color maps.  Therefore, we were able
to add the $g'$, $r'$ and  $i'$ images corresponding to the same field
in  order to  improve the  S/N ratio  by enhancing  the extremely  low
surface  brightnesses  of these  galaxies,  thus  allowing ELLIPSE  to
converge  with as  many  free parameters  as  possible.  The  isophote
tables obtained from the added images  were used as a reference to get
the tables corresponding to each individual filter.

The instrumental  brightness profiles  were corrected by  the residual
sky  level through  the analysis  of the  curves of  growth (COG;  for
details,    we     refer    the    reader    to     section\,3.1    of
\citealt{2016A&A...596A..23S})  and,  later,  calibrated to  the  SDSS
photometric system.  To do so,  we used standard star  fields observed
within our programs. The expression used to calibrate our data was:
\begin{equation}
  \mu_\mathrm{std}=\mu_\mathrm{zero}+\mu_\mathrm{inst}+2.5~\log(sca^2~t_\mathrm{exp})
  -K~(X-1.0)\textrm{,}
\end{equation}
where  $\mu_\mathrm{std}$  is  the  surface  brightness  (in  standard
magnitudes  arcsec$^{-2}$),  $\mu_\mathrm{zero}$  is  the  photometric
zero-point,   $\mu_\mathrm{inst}$   is    the   instrumental   surface
brightness, $sca$ is the scale  of the images, $t_\mathrm{exp}$ is the
exposure time,  $K$ is the  mean atmospheric extinction at  Mauna Kea,
and $X$ is the airmass value.

\begin{figure}[th!]
\center
\hspace{0.55cm}
\includegraphics[scale=0.25]{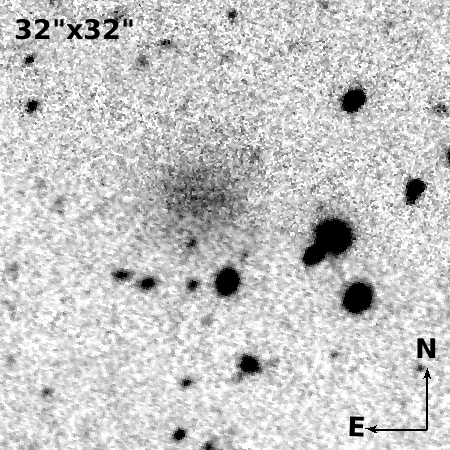}
\includegraphics[scale=0.25]{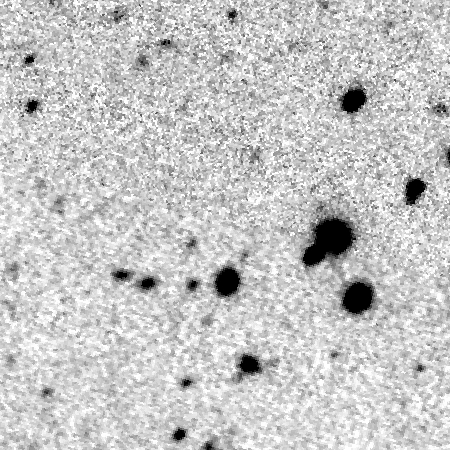}
\resizebox{1.0\hsize}{!}{\includegraphics{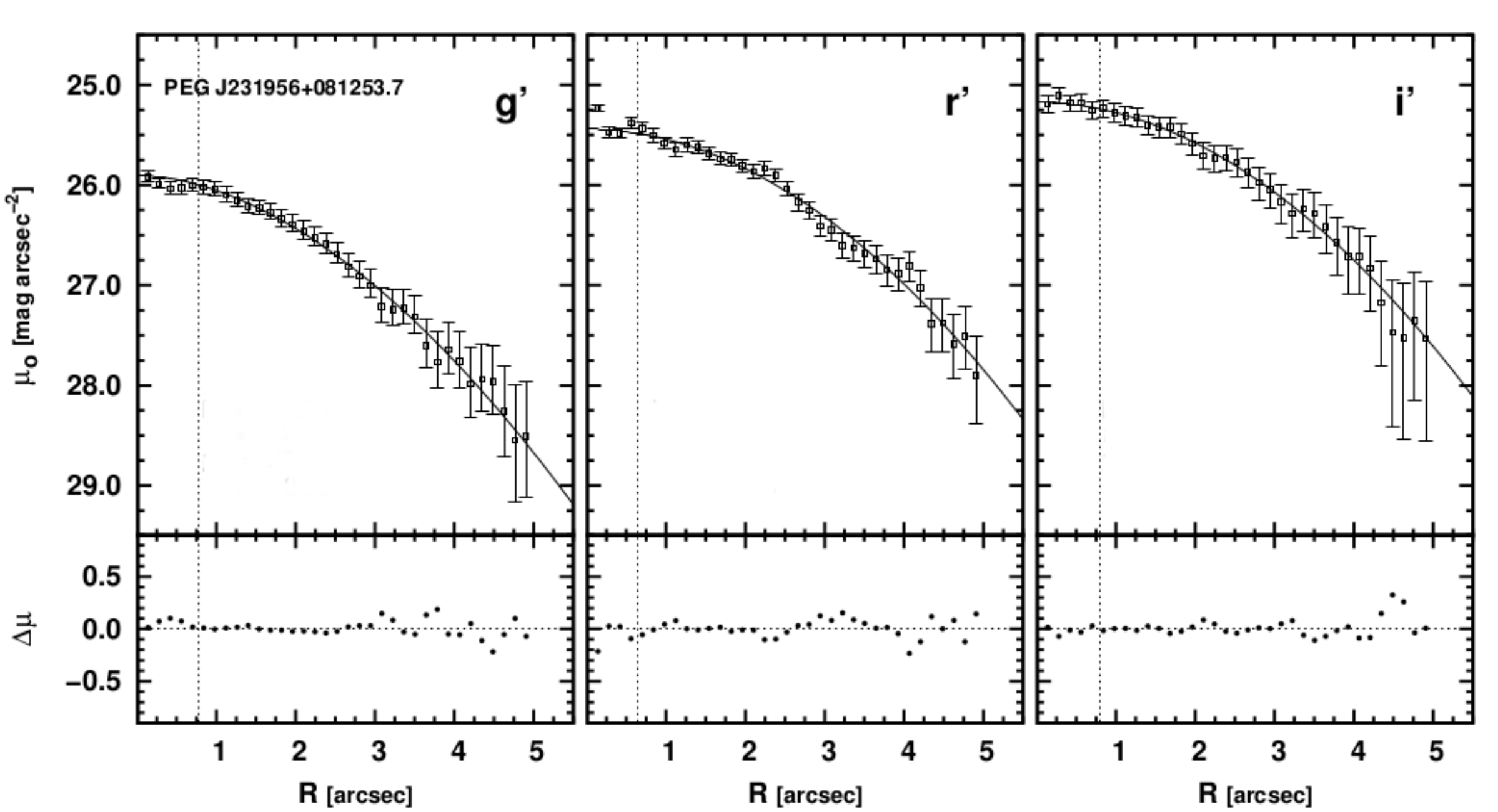}}
\caption{{\it Upper panels:}  32 $\times$ 32 arcsec frame  in the $r'$
  filter showing the LSB galaxy PEG\,J231956+081253.7 {\it (left)} and
  the residuals  left by the subtraction  of the model built  from the
  brightness profile obtained with  ELLIPSE {\it (right)}.  {\it Lower
    panels:} S\'ersic law fits to  the absorption corrected $g'$, $r'$
  and    $i'$    brightness    profiles     of    the    LSB    galaxy
  PEG\,J231956+081253.7.   The  residuals   $\Delta\mu  =  \mu(\mathrm{obs})  -
  \mu(\mathrm{fit})$ are shown  in the lower panel.  The  dotted vertical lines
  indicate the  inner region of  the profiles excluded to  perform the
  fits in order to avoid seeing effects.
}
\label{J231956+081253}
\end{figure}

The upper  panels of Figure\,\ref{J231956+081253} show  the $g'$ image
of one  of the LSB galaxies  detected in Pegasus\,I ({\it  left}), and
the  residuals  left  by  the  model  subtraction  ({\it  right};  see
Sect.~\ref{sersic_profiles}). Similar figures for the rest of Pegasus'
LSB galaxies can be found in Section\,\ref{apendice}. 

In  Table\,\ref{COG}  we  present  the  model-independent  photometric
parameters obtained from  the $g'$, $r'$ and  $i'$ observed brightness
profiles.  In  this case,  total magnitudes  are obtained  through the
numerical integration  of these  profiles up  to the  total equivalent
radius  ($r_\mathrm{tot}$)  of  the galaxy  (see  Table\,\ref{COG}).  We
consider  that $r_\mathrm{tot}$ is  the equivalent  radius at  which the
curve of  growth stabilizes in  the three filters.   The corresponding
errors were calculated from the integration of the brightness profiles
obtained after adding and subtracting the median value of the residual
sky level to the original profile.
\subsection{S\'ersic profiles}
\label{sersic_profiles}

\begin{table*}[h!]
  \centering
\scriptsize
\caption{Structural  parameters  of  the   LSB  galaxy  candidates  in
  Pegasus\,I,  obtained  from  the  fit  of  S\'ersic  laws  to  their
  brightness  profiles. $r_\mathrm{int}$  and $r_\mathrm{ext}$  denote
  the internal and  external radii considered to perform  the fits. In
  the  case  of the  last  object  in  the  list, which  displays  two
  components in its brightness profile,  a combination of two S\'ersic
  functions was fitted to the entire profile {\it (see text)}. }
\begin{tabular}{lcccccccccccccc}
\hline 
\hline
Object & \multicolumn{1}{c}{Filter} &\multicolumn{1}{c}{$r_{\rm int}$} &\multicolumn{1}{c}{$r_{ext}$}   & \multicolumn{1}{c}{$\mu_{0}$}           & \multicolumn{1}{c}{$r_{0}$}  & \multicolumn{1}{c}{N} & \multicolumn{1}{c}{${\mu}_{{\rm eff}}$}    & \multicolumn{1}{c}{${r^s}_{\rm eff}$}   &  \multicolumn{1}{c}{$n$} &  \multicolumn{1}{c}{${m^s}_{\rm tot}$}\\
       &                            &\multicolumn{1}{c}{(arcsec)} &\multicolumn{1}{c}{(arcsec)}   & \multicolumn{1}{c}{(mag~arcsec$^{-2}$)} & \multicolumn{1}{c}{(arcsec)} &                      & \multicolumn{1}{c}{(mag~arcsec$^{-2}$)} & \multicolumn{1}{c}{(arcsec)}   &          &  \multicolumn{1}{c}{(mag)}                  \\
(1)    & (2)                        & (3)                         & (4)                           & (5)                                     & (6)                          & (7)                  & (8)                                     & (9)                           & (10)          & (11)     \\  
\hline
\hline
                      & $g'$ & 0.8 &    & 25.90 $\pm$ 0.03 & 2.97 $\pm$ 0.05 & 1.80 $\pm$ 0.07  & 26.75 & 2.60 & 0.56 & 22.23 \\
PEG J231956+081253.7  & $r'$ & 0.7 & 5  & 25.44 $\pm$ 0.03 & 3.34 $\pm$ 0.07 & 1.97 $\pm$ 0.13  & 26.19 & 2.77 & 0.51 & 21.57 \\
                      & $i'$ & 0.8 &    & 25.17 $\pm$ 0.02 & 3.30 $\pm$ 0.05 & 1.95 $\pm$ 0.09  & 25.93 & 2.74 & 0.51 & 21.33 \\
\hline                                             
                      & $g'$ & 0.8 &    & 26.82 $\pm$ 0.05 & 8.60  $\pm$ 0.35 & 1.36 $\pm$ 0.10  & 28.05 & 9.46 & 0.73 & 20.62 \\
PEG J232023+081331.4  & $r'$ & 0.8 & 15 & 26.18 $\pm$ 0.03 & 10.61 $\pm$ 0.24 & 1.81 $\pm$ 0.12  & 27.03 & 9.24 & 0.55 & 19.76 \\
                      & $i'$ & 0.9 &    & 25.73 $\pm$ 0.02 & 9.93  $\pm$ 0.17 & 1.88 $\pm$ 0.09  & 26.53 & 8.43 & 0.53 & 19.48 \\
\hline                                              
                      & $g'$ & 0.9 &    & 26.27 $\pm$ 0.02& 11.24 $\pm$ 0.16 & 1.63 $\pm$ 0.08  & 27.25 & 10.52 & 0.64 & 19.69 \\
PEG J232024+081209.0  & $r'$ & 1.0 & 14 & 25.51 $\pm$ 0.02& 10.04 $\pm$ 0.16 & 1.69 $\pm$ 0.08  & 26.44 & 9.16  & 0.59 & 19.19 \\
                      & $i'$ & 0.7 &    & 25.64 $\pm$ 0.03& 10.35 $\pm$ 0.27 & 1.74 $\pm$ 0.14  & 26.53 & 9.25  & 0.57 & 19.31 \\
\hline                                                                                           
                      & $g'$ & 1.0 &    & 25.28 $\pm$ 0.02 & 5.30 $\pm$ 0.09 & 1.37 $\pm$ 0.04 & 26.51 & 5.81  & 0.73 & 20.13 \\
PEG J232037+080934.3  & $r'$ & 0.9 & 10 & 24.80 $\pm$ 0.02 & 4.93 $\pm$ 0.10 & 1.36 $\pm$ 0.04 & 26.04 & 5.42  & 0.73 & 19.82 \\
                      & $i'$ & 0.9 &    & 24.51 $\pm$ 0.02 & 4.89 $\pm$ 0.10 & 1.33 $\pm$ 0.04 & 25.79 & 5.52  & 0.75 & 19.51 \\
\hline                                                                                                    
                      & $g'$ & 0.8 &    & 25.55 $\pm$ 0.03 & 2.78 $\pm$ 0.05 & 1.92 $\pm$ 0.09 & 26.32 & 2.33  & 0.52 & 22.06 \\
PEG J232037+081336.6  & $r'$ & 0.8 & 4  & 25.08 $\pm$ 0.02 & 2.98 $\pm$ 0.03 & 2.26 $\pm$ 0.08 & 25.69 & 2.30  & 0.44 & 21.48 \\
                      & $i'$ & 0.9 &    & 24.65 $\pm$ 0.03 & 2.84 $\pm$ 0.06 & 1.62 $\pm$ 0.09 & 25.63 & 2.68  & 0.62 & 21.03 \\
\hline 
                      & $g'$ & 0.8 &    & 24.95 $\pm$ 0.05 & 2.77 $\pm$ 0.10 & 1.43 $\pm$ 0.07 & 26.11 & 2.91  & 0.70 & 21.20 \\
PEG J232038+081046.9  & $r'$ & 0.8 & 6  & 24.42 $\pm$ 0.04 & 2.53 $\pm$ 0.08 & 1.25 $\pm$ 0.05 & 25.80 & 3.07  & 0.80 & 20.74 \\
                      & $i'$ & 0.9 &    & 24.13 $\pm$ 0.06 & 2.58 $\pm$ 0.13 & 1.19 $\pm$ 0.07 & 25.60 & 3.34  & 0.84 & 20.30 \\
\hline
                      & $g'$ & 0.6 &    & 25.96 $\pm$ 0.03 & 3.59 $\pm$ 0.11 & 1.26 $\pm$ 0.07 & 27.32 & 4.30  & 0.79 & 21.63 \\
PEG J232049+080806.2  & $r'$ & 0.6 & 6  & 25.50 $\pm$ 0.03 & 3.64 $\pm$ 0.09 & 1.52 $\pm$ 0.08 & 26.57 & 3.62  & 0.66 & 21.47 \\
                      & $i'$ & 0.5 &    & 24.94 $\pm$ 0.03 & 3.35 $\pm$ 0.17 & 1.03 $\pm$ 0.07 & 26.69 & 5.36  & 0.97 & 20.66 \\
\hline
                      			& $g'$ &     &    & 24.32 $\pm$ 0.01 & 0.45 $\pm$ 0.01 & 2.11 $\pm$ 0.11 & 24.99 & 0.36  & 0.47 & 24.83 \\
PEG J232054+080838.8 {\it (int)} 	& $r'$ & 0.0 &6.5 & 23.46 $\pm$ 0.01 & 0.44 $\pm$ 0.01 & 2.18 $\pm$ 0.10 & 24.10 & 0.34  & 0.46 & 24.07 \\
                      			& $i'$ &     &    & 22.68 $\pm$ 0.01 & 0.33 $\pm$ 0.01 & 1.89 $\pm$ 0.06 & 23.48 & 0.28  & 0.53 & 23.82 \\
\hline
                      			& $g'$ &     &    & 25.13 $\pm$ 0.01 & 1.86 $\pm$ 0.30 & 0.80 $\pm$ 0.08 & 27.48 & 4.88  & 1.25 & 21.24 \\
PEG J232054+080838.8 {\it (out)}  	& $r'$ & 0.0 &6.5 & 24.89 $\pm$ 0.01 & 2.48 $\pm$ 0.44 & 0.92 $\pm$ 0.13 & 26.89 & 5.74  & 1.08 & 20.37 \\
                      			& $i'$ &     &    & 24.50 $\pm$ 0.02 & 2.14 $\pm$ 0.48 & 0.80 $\pm$ 0.12 & 26.87 & 5.67  & 1.29 & 20.30 \\
\hline
\hline
\end{tabular}
\label{sersic}
\end{table*}
 
Due to the extremely low surface brightnesses of our objects, we would
like to compare our  model-independent parameters, with those obtained
from  fitting a  general S\'ersic  law \citep{1963BAAA....6...41S}  to
their surface brightness profiles:
\begin{equation}
\label{eq1}
\mu (r)=\mu_{0}+1.0857~\left(\dfrac{r}{r_{0}} \right)^{N}\textrm{.}
\end{equation}
\noindent  Here,  $\mu_{0}$  designates the  central  ($r=0$)  surface
brightness, $r_{0}$ is  the scalelength of the profile and  $N$ is the
S\'ersic index. Due to its simpler mathematical dependence on the free
parameters, we decided to use the above formula instead of:
\begin{equation}
\mu(r)=\mu_\mathrm{eff}+1.0857~b_n\left[\left(\dfrac{r}{{r^s}_\mathrm{eff}}
  \right)^{\frac{1}{n}} - 1\right]\textrm{,}
\end{equation}
\noindent  where  $b_{n}~\simeq~1.9992~n-0.3271$  for $0.5  <  n<  10$
\citep[][and   references    therein]{2008MNRAS.388.1708G}   and   the
superscript  $s$ refers  to  quantities obtained  assuming a  S\'ersic
profile.  There  are simple relations between  the quantities involved
in both equations \citep[e.g.][]{2003ApJ...582..689M}:
\begin{equation}
  n=\dfrac{1}{N}
\end{equation}
\begin{equation}
  \mu_\mathrm{eff}=\mu_0+1.0857~b_n
\end{equation}
\begin{equation}
 {r^s}_\mathrm{eff}=r_0~b_n^n.
\end{equation}

All        galaxies,       excepting        PEG       J232054+080838.8
(Figure\,\ref{J232054+080838}),  were fitted  with  a single  S\'ersic
law, excluding  the inner regions  affected by seeing.  In  each case,
the fit extended from an inner radius defined by the seeing FWHM, to a
maximum  radius set  at  the  point where  the  profile  begins to  be
significantly affected by small variations in the sky level.

PEG J232054+080838.8 displays what seems to  be a nucleus.  As a first
step,  we built  a PSF  model of  this star-like  object with  DAOPHOT
within IRAF.  After subtracting this  model to the original  images of
the  galaxy,   the  $g'$,  $r'$,   and  $i'$  frames   displayed  some
residuals.  This  was  taken  as  an  evidence  that  the  nucleus  is
marginally resolved in  all the images.  Therefore, we  decided to fit
the sum of two S\'ersic functions to the whole g', r' and i' profiles,
without  excluding the  very inner  region with  the aim  at obtaining
structural parameters for the central component. We tested the results
of the  composite fit considering,  first, an external  component with
$n=1$ and,  then, a fit with  free parameters for both  components. We
have found that  allowing all the parameters to vary  freely gives the
minimum $\chi^2$.

In Table\,\ref{sersic} we present the resulting structural parameters,
as well as the internal and  external radius considered to perform the
fits in  each case (columns  4 and 5, respectively).  Total integrated
magnitudes were also obtained from the expression:
\begin{multline}
  {m^s}_{\rm tot}=\mu_\mathrm{eff}-1.995450-5~\log({r^s}_\mathrm{eff})\\
  -1.0857~b_n-2.5~\log\left[b_n^{-2n}~n\Gamma(2n)\right].
\end{multline}

From  Table\,\ref{COG} and  Table\,\ref{sersic}, it  can be  seen that
S\'ersic  magnitudes   and  $r_{\rm   eff}$  are  in   agreement  with
model-independent ones, within the errors. As an example, in the lower
panels  of  Figure\,\ref{J231956+081253},  we  show the  fits  to  the
brightness profiles, and their residuals,  for one of the LSB galaxies
detected in Pegasus\,I.  Similar figures  for the rest of Pegasus' LSB
galaxies are shown in Section\,\ref{apendice}.
\section{Comparison with other samples}
\label{Samples}
\subsection{Color-magnitude diagram}
\label{CMD}
\begin{figure*}[th!]
  \center
\resizebox{\hsize}{!}{\includegraphics{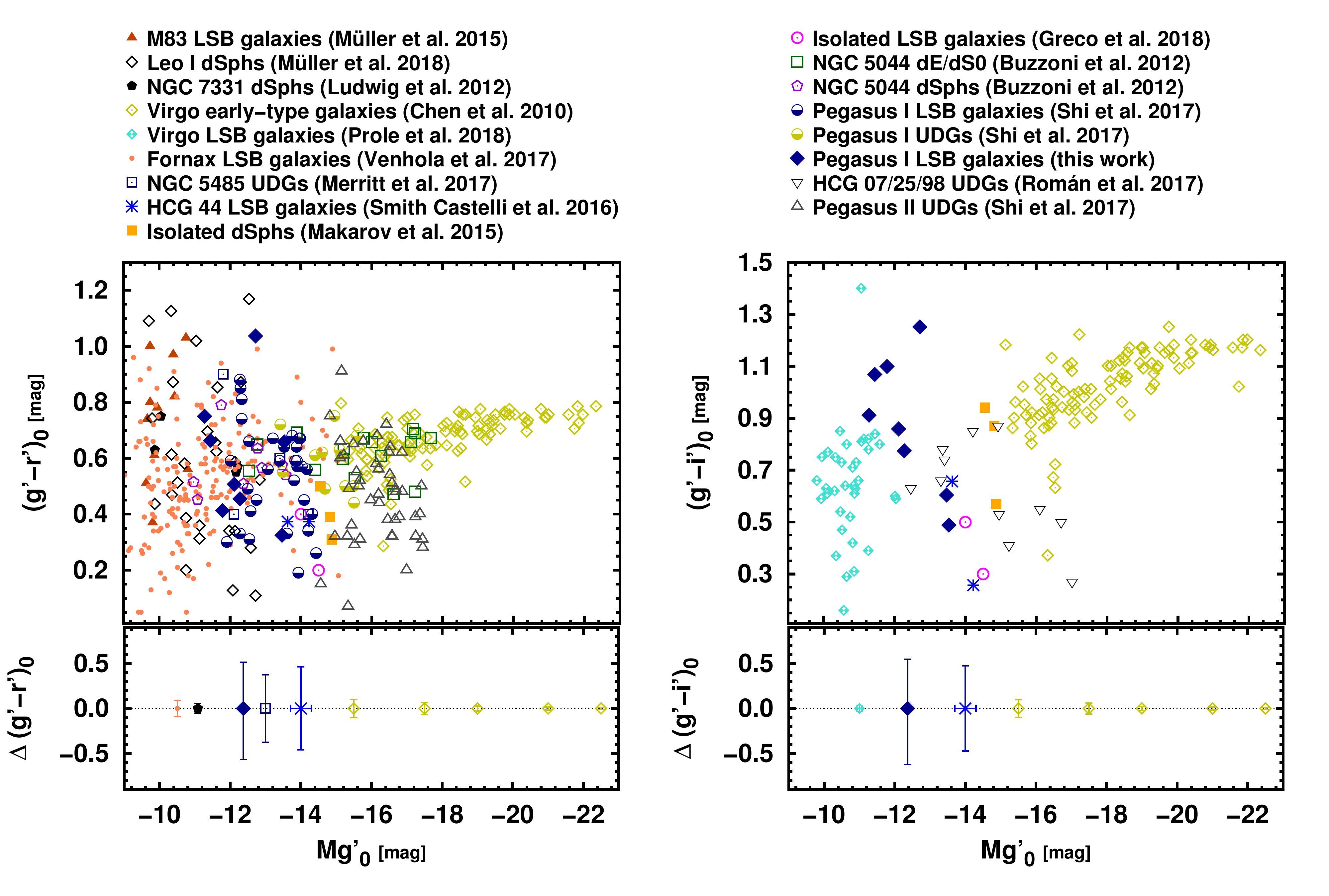}}
\caption{Color-magnitude  diagrams showing  the  location  of the  LSB
  galaxies  presented in  this work,  assuming  that they  are at  the
  distance of Pegasus\,I ({\it blue filled diamonds}). As a reference,
  we  plot the  red  sequence  defined by  a  subsample of  early-type
  galaxies in the central region of the Virgo cluster ({\it green open
    diamonds}; \citealp{2010ApJS..191....1C}), and  a sample of dE/dS0
  galaxies in the  NGC\,5044 group ({\it open squares  without a dot};
  \citealp{2012MNRAS.420.3427B}).  We  also include  different samples
  of LSB,  dwarf spheroidal (dSph), and  ultra-diffuse galaxies (UDGs)
  reported in the literature. They are listed in the top margin of the
  plot,  according to  their distances  or those  of the  environments
  towards they were  detected, in ascending order from  top to bottom,
  and from left to right. They are:
  {\it M\,83 LSB galaxies} \citep{2015A&A...583A..79M},
  {\it Leo\,I  dSphs} \citep{2015A&A...583A..79M},
  {\it NGC\,7331 dSphs} \citep{2012AJ....144..190L},
  {\it Virgo LSB galaxies} \citep{2015ApJ...809L..21M,2018MNRAS.478..667P},
  {\it Fornax LSB galaxies} \citep{2017A&A...608A.142V},
  {\it NGC\,5485 UDGs} \citep{2017ApJ...846...26S},
  {\it HCG\,44 LSB galaxies} \citep{2016A&A...596A..23S},
  {\it Isolated dSphs} \citep{2015A&A...581A..82M},
  {\it Isolated LSB galaxies} \citep{2018arXiv180504118G},
  {\it NGC\,5044 dSphs} \citep{2012MNRAS.420.3427B},
  {\it Pegasus\,I LSB/UDGs} \citep{2017ApJ...846...26S},
  {\it HCG\,07, HCG\,25 and HCG\,98} \citep{2017MNRAS.468.4039R} and
  {\it Pegasus\,II UDGs} \citep{2017ApJ...846...26S}.
  In  both diagrams,  the mean  error bars  of the  different samples,
  when available, are shown in the panels below.}
\label{CMDs}
\end{figure*}
It is  interesting to  compare the photometric  properties of  the LSB
galaxies detected in the direction of Pegasus\,I with those of similar
objects   identified  in   other  environments.    To  this   aim,  in
Figure\,\ref{CMDs}  we  present  two  color-magnitude  diagrams  (CMD)
showing the location of dwarf spheroidal (dSph), LSB and ultra-diffuse
galaxies  (UDGs)  reported  in  the literature,  along  with  the  LSB
galaxies analyzed  in this work, assuming  that the latter are  at the
distance of Pegasus\,I. We consider the distance modulus of Pegasus\,I
as that obtained from the mean  distance of the two dominant galaxies,
NGC\,7619   and   NGC\,7626   (Table\,\ref{distancias};   $\langle   D
\rangle=50.8$\,Mpc, $(m-M)=33.52$\,mag, 1 arcsec  = 0.248\,kpc).  As a
reference, we  also show the  red sequence  defined by a  subsample of
early-type galaxies in the Virgo cluster \citep{2010ApJS..191....1C}.

We can  see from these  plots that, in  general, LSB galaxies  seem to
display a much wider range  in color than typical early-type galaxies.
In the  particular case of  the LSB objects identified  in Pegasus\,I,
with the  exception of one  case, they tend  either to follow  the red
sequence of early-type galaxies or to show redder colors.  Remarkably,
the reddest LSB galaxies of the  considered samples are found in quite
rich  environments  like M\,83,  Leo\,I  and  Pegasus\,I, while  their
colors  are  much bluer  than  those  expected  for Milky  Way  cirrus
($(g'-r')=1.33$ to 2.03; \citealp{2012AJ....144..190L}).  In contrast,
LSB  galaxies found  in isolation  or within  Hickson compact  groups,
either follow the red sequence of early-type galaxies or display bluer
colors.  The LSB  galaxies reported  in  the Virgo  cluster, with  the
exception of one object, also behave in the same manner.
\subsection{$\langle\mu_{\rm eff}\rangle$-luminosity diagram}
\begin{figure*}[t!]
\center
\resizebox{\hsize}{!}{\includegraphics[scale=1.5]{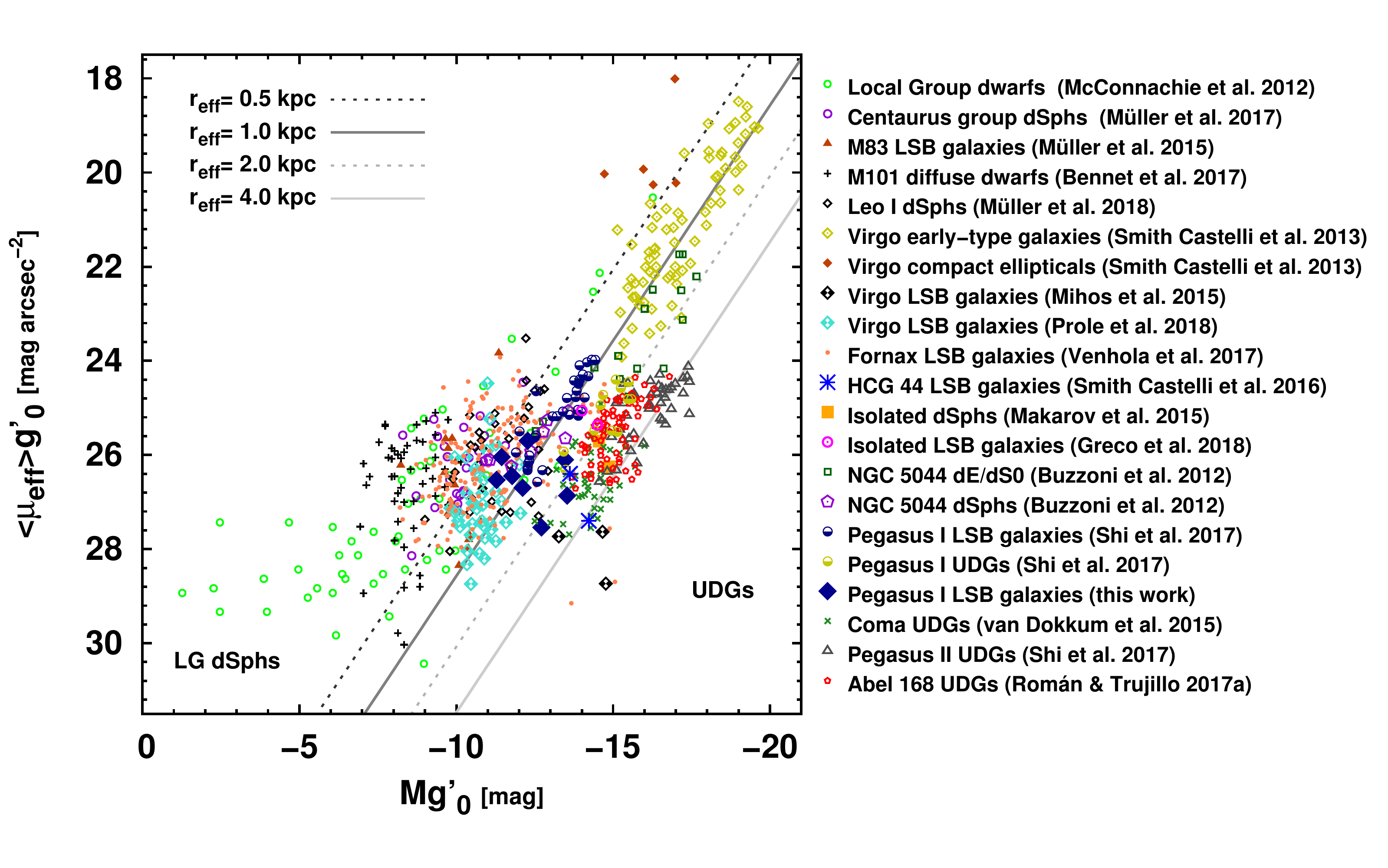}}
\caption{$\langle \mu_\mathrm{eff} \rangle$-luminosity  diagram of the
  early-type  galaxies in  the  central region  of  the Virgo  cluster
  \citep{2013ApJ...772...68S},  showing   the  location  of   the  LSB
  galaxies presented in  this work, assuming they are  at the distance
  of Pegasus\,I. For comparison, we  also include different samples of
  LSB,  dwarf spheroidal  (dSph),  and  ultra-diffuse galaxies  (UDGs)
  reported in the  literature. They are listed in the  right margin of
  the plot according  to their distances or those  of the environments
  towards  they  were  detected,  in   ascending  order  from  top  to
  bottom. They are:
  {\it Local Group dwarfs} \citep{2012AJ....144....4M},
  {\it Centaurus dSphs} \citep{2017A&A...597A...7M},
  {\it M\,83 LSB galaxies} \citep{2015A&A...583A..79M},
  {\it M\,101 dwarfs} \citep{2017ApJ...850..109B},
  {\it Leo\,I dSphs} \citep{2015A&A...583A..79M}, 
  {\it Virgo LSB galaxies} \citep{2015ApJ...809L..21M,2018MNRAS.478..667P},
  {\it Fornax LSB galaxies} \citep{2017A&A...608A.142V},
  {\it HCG\,44 LSB galaxies} \citep{2016A&A...596A..23S},
  {\it Isolated dSphs} \citep{2015A&A...581A..82M},
  {\it Isolated LSB galaxies} \citep{2018arXiv180504118G},
  {\it NGC\,5044 dE/dS0 and dSph galaxies} \citep{2012MNRAS.420.3427B},
  {\it Pegasus\,I LSB/UDGs} \citep{2017ApJ...846...26S},
  {\it Coma UDGs} \citep{2015ApJ...798L..45V},
  {\it Pegasus\,II UDGs} \citep{2017ApJ...846...26S} and 
  {\it Abell\,168 UDGs} \citep{2017MNRAS.468..703R}.
  The lines of constant $r_\mathrm{eff}$ are only valid for exponential profiles.}
\label{mueff}
\end{figure*}

In  Figure\,\ref{mueff}   we  present   the  $\langle   \mu_{\rm  eff}
\rangle$--luminosity  diagram of  different  samples  of LSB  galaxies
reported in the  literature.  As a reference, we show  the location of
subsamples   of    early-type   galaxies   in   the    Virgo   cluster
\citep{2013ApJ...772...68S}      and      the     NGC\,5044      group
\citep{2012MNRAS.420.3427B}.

This plot shows that, as already reported, typical early-type galaxies
of  different luminosities  tend  to  be placed  around  the locus  of
constant $r_\mathrm{eff}=1$\,kpc with a  low dispersion, regardless of
their         environment          \citep[e.g.][and         references
  therein]{2012MNRAS.419.2472S,2013ApJ...772...68S}.  In contrast, LSB
galaxies tend to  display a wider range  in $r_\mathrm{eff}$, spanning
from LG dSph  galaxies (the smallest LSB galaxies detected  so far) to
the so-called UDGs in more distant  groups and clusters (which are the
largest examples  of LSB  galaxies). In  particular, the  LSB galaxies
detected in Pegasus\,I seem to  show $r_\mathrm{eff}$ similar to those
of early-type galaxies,  filling the gap between the  smallest and the
largest LSB systems.
\begin{figure*}[th!]
\center
\resizebox{\hsize}{!}{\includegraphics[scale=1.7]{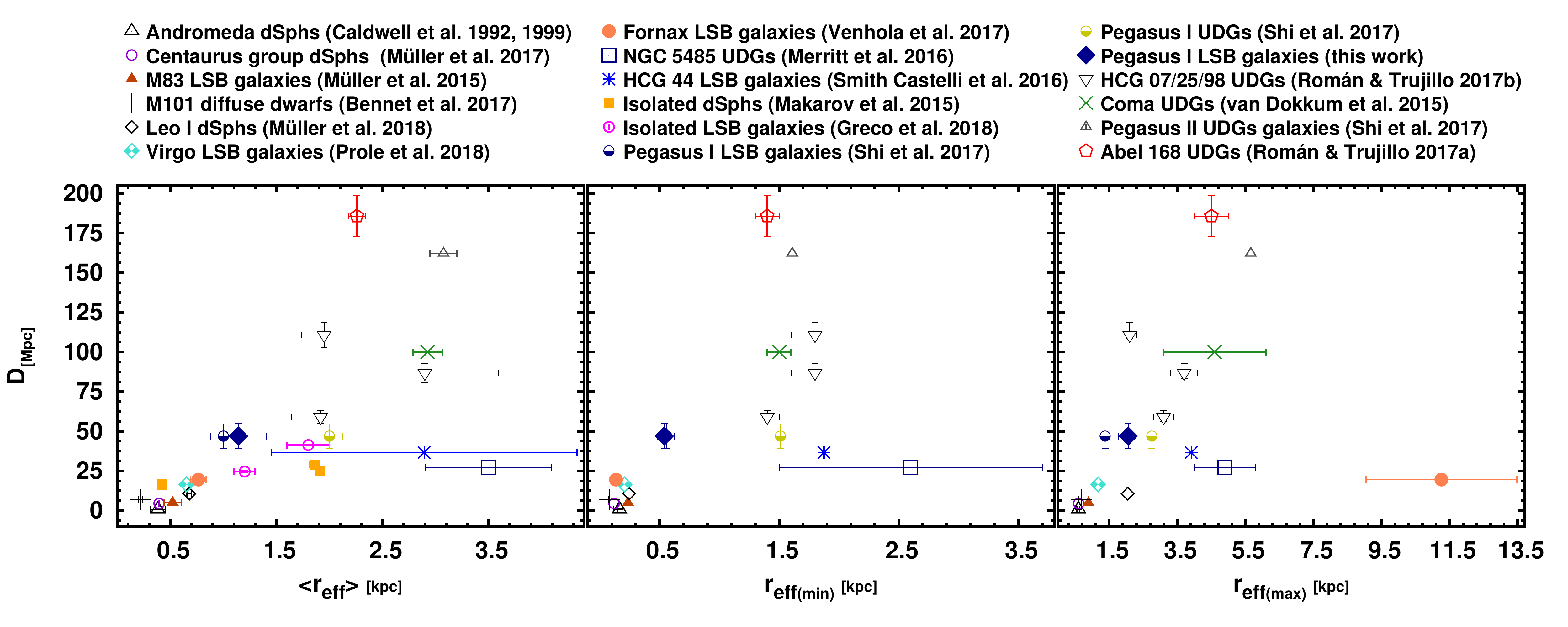}}
\resizebox{\hsize}{!}{\includegraphics[scale=1.7]{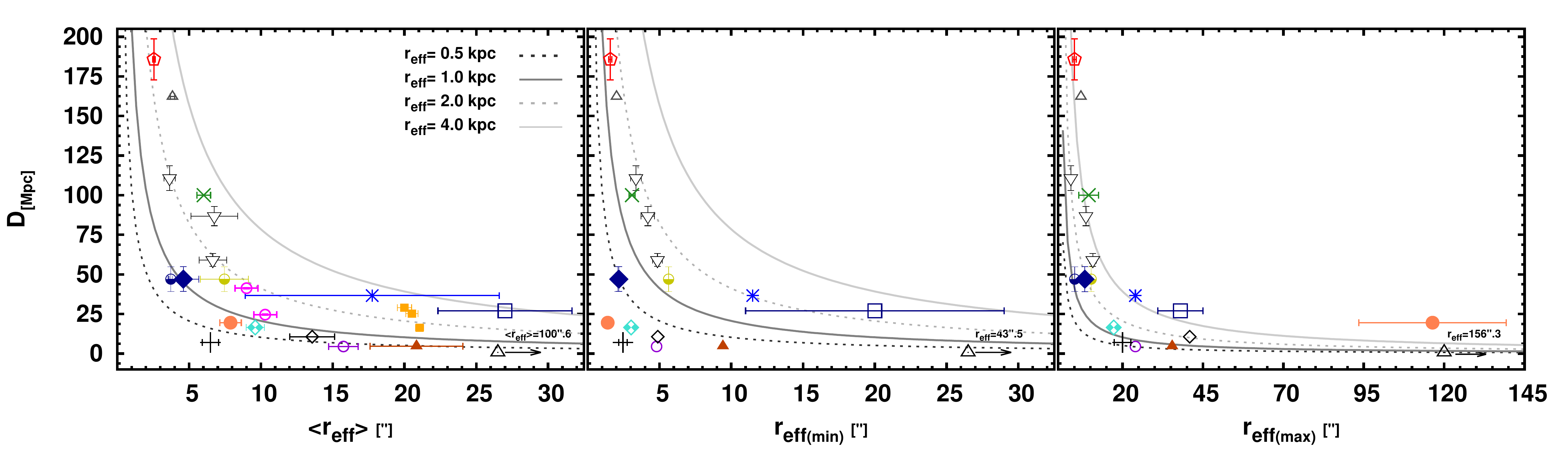}}
\caption{Mean {\it  (left)}, minimum  {\it (middle)} and  maximum {\it
    (right)}  absolute   {\it  (top)}  and  apparent   {\it  (bottom)}
  effective  radius of  several  samples of  LSB  galaxies versus  the
  distances to  the environments towards  they were detected.   In the
  upper  panels a  trend  is  evident in  the  sense  that larger  LSB
  galaxies  are  detected  in  the most  distant  environments,  while
  smaller LSB  galaxies are missed.   The curves in the  bottom panels
  are only valid for exponential profiles.}
\label{Distances}
\end{figure*}

In  this plot,  the references  to the  different samples  are ordered
according to  their reported  distances (or  those of  the environments
towards  they  were  detected),  increasing  from  top  to  bottom.   A
distance-size trend  is apparent;  that is,  the smallest  objects are
identified in the  Local Group, while the largest  systems reported in
the literature are detected in more distant groups/clusters.

To quantify this appreciation,  Figure\,\ref{Distances} shows the mean,
the minimum and  the maximum (absolute and apparent)  $r_\mathrm{eff}$ of
different  samples.  All  the panels  show the  same trend:  while the
largest objects are  found in distant environments,  the smallest ones
are only detected in the Local Volume.
\subsection{Andromeda's satellites}
\label{Andromeda}

\begin{figure}[th!]
\centering
\resizebox{\hsize}{!}{\includegraphics{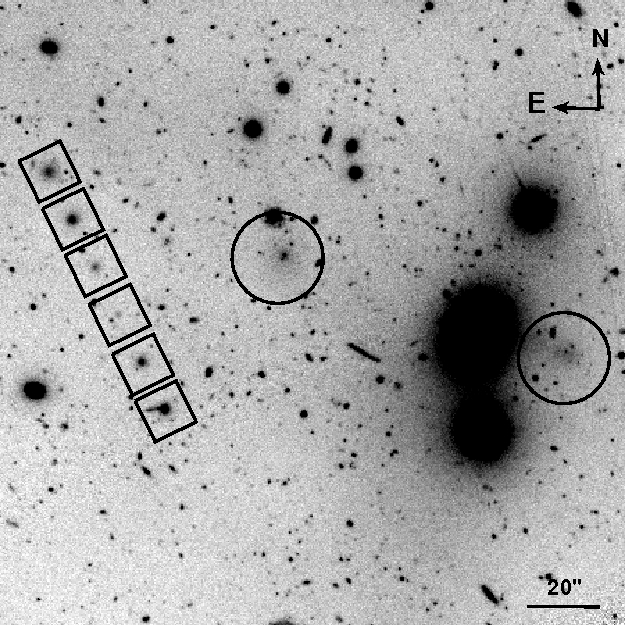}}
\caption{2.85 $\times$  2.85 arcmin frame  in the $g'$  filter showing
  two of the  LSB galaxies detected in Pegasus\,I  {\it (circles)}. In
  order  to compare  the appearance  that dSph  Andromeda's satellites
  would have at the distance of Pegasus\,I, we added to this image the
  models     obtained      from     their      brightness     profiles
  \citep{1992AJ....103..840C,1999AJ....118.1230C}   scaled   to   that
  distance {\it (squares)}.}
\label{SAnd_Peg}  
\end{figure}

\begin{figure*}[th!]
\center
\hspace{0.9cm}
\resizebox{1.0\hsize}{!}{\includegraphics{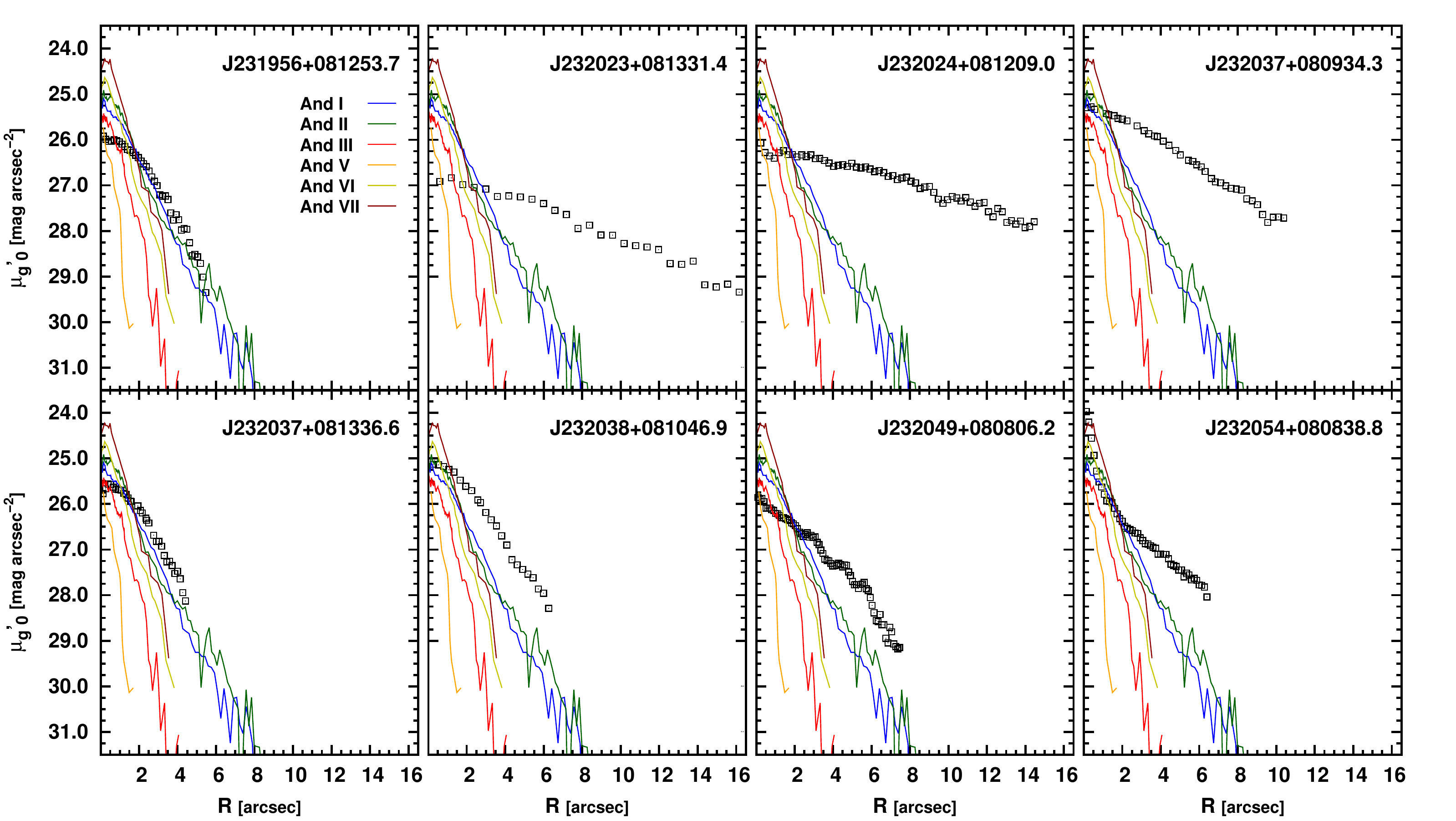}}
\caption{Comparison between  the $g'$  brightness profiles of  the LSB
  galaxies  detected in  Pegasus\,I ({\it  black open  squares with  a
    dot)}    and     those    of     six    M31     dSph    satellites
  \citep{1992AJ....103..840C,1999AJ....118.1230C}.         Andromeda's
  satellites  profiles  were  transformed  from the  $V$-band  to  the
  $g'$-band      through      the     relations      presented      by
  \citet{1995PASP..107..945F},   and  scaled   to   the  distance   of
  Pegasus\,I.}
\label{perfiles-SAnd_Peg}
\end{figure*}

Regarding  the likely  presence  of a  bias in  the  detection of  LSB
galaxies, we wonder which would be the appearance of, for example, the
M31  (Andromeda)  dSph  satellites  at  the  distance  of  Pegasus\,I.
Figure\,\ref{SAnd_Peg} shows models of  six of these galaxies obtained
from the  brightness profiles reported  by \citet{1992AJ....103..840C}
and \citet{1999AJ....118.1230C}, overimposed on one of our Gemini-GMOS
fields.   In addition,  in Figure\,\ref{perfiles-SAnd_Peg}  we compare
the surface brightness profiles of our  LSB galaxies with those of the
Andromeda's  satellites scaled  to the  distance of  Pegasus\,I.  From
these  figures it  can be  seen that  Andromeda's dSphs  would display
$2<r_\mathrm{tot}<8$ arcsec, and  $24<\mu_0{_g}<26$ mag arcsec$^{-2}$.
Their central surface brightnesses would  thus be quite low, but still
brighter  than  those of  the  LSB  galaxies detected  in  Pegasus\,I;
however, their apparent sizes would be much smaller in comparison with
the LSB galaxies in our sample.

 Therefore, it is  evident that we have certainly  missed any putative
 objects similar  to the  dSph satellites of  Andromeda in  our visual
 selection of extended and diffuse  galaxies.  In a following paper we
 will present  the analysis of  our images regarding the  detection of
 counterparts  of the  Andromeda's  dSph  galaxies towards  Pegasus\,I
 (Gonz\'alez et al., in preparation).
\subsection{S\'ersic index}

\begin{figure}[h!]
\centering
\resizebox{\hsize}{!}{\includegraphics{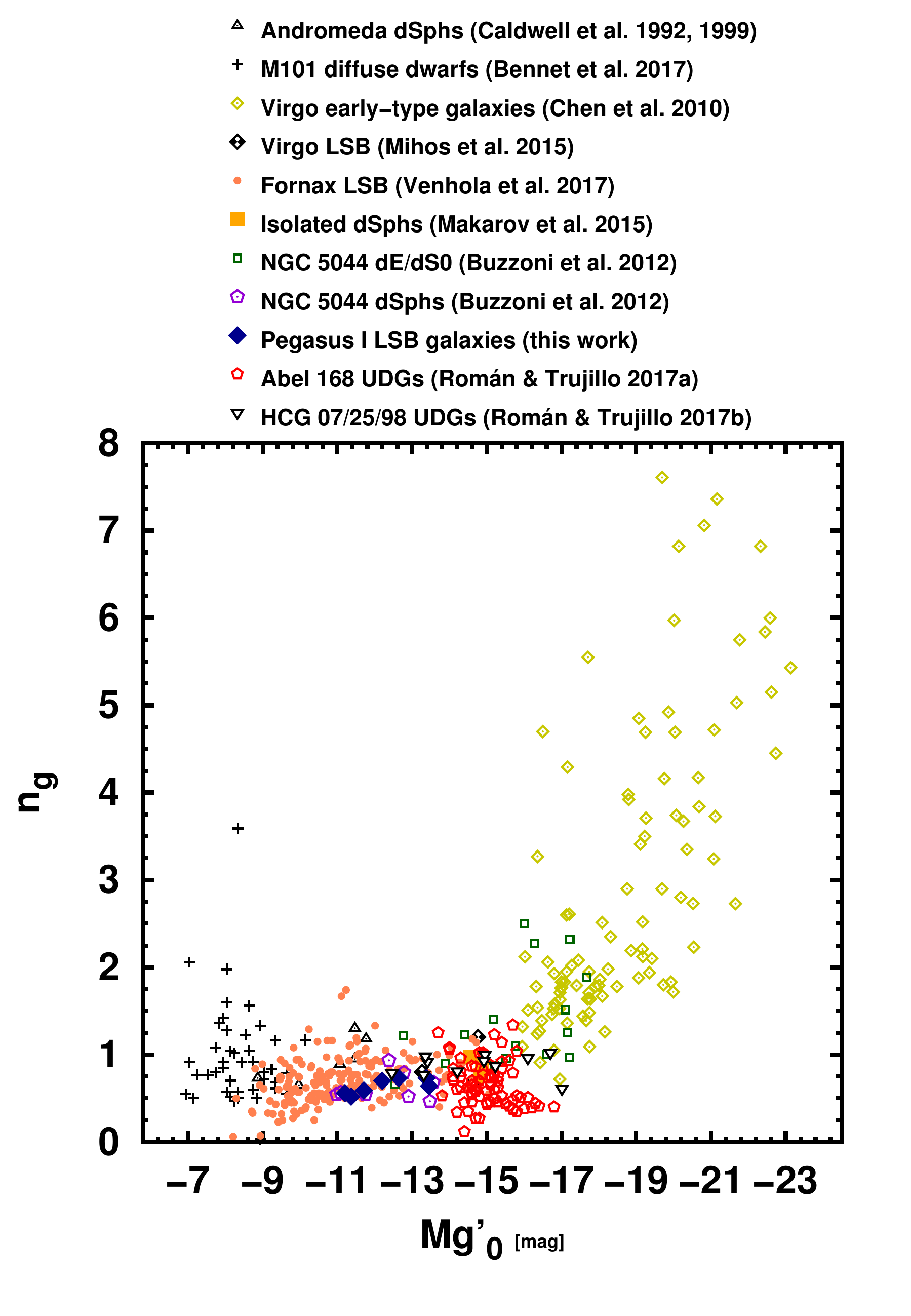}}
\resizebox{\hsize}{!}{\includegraphics{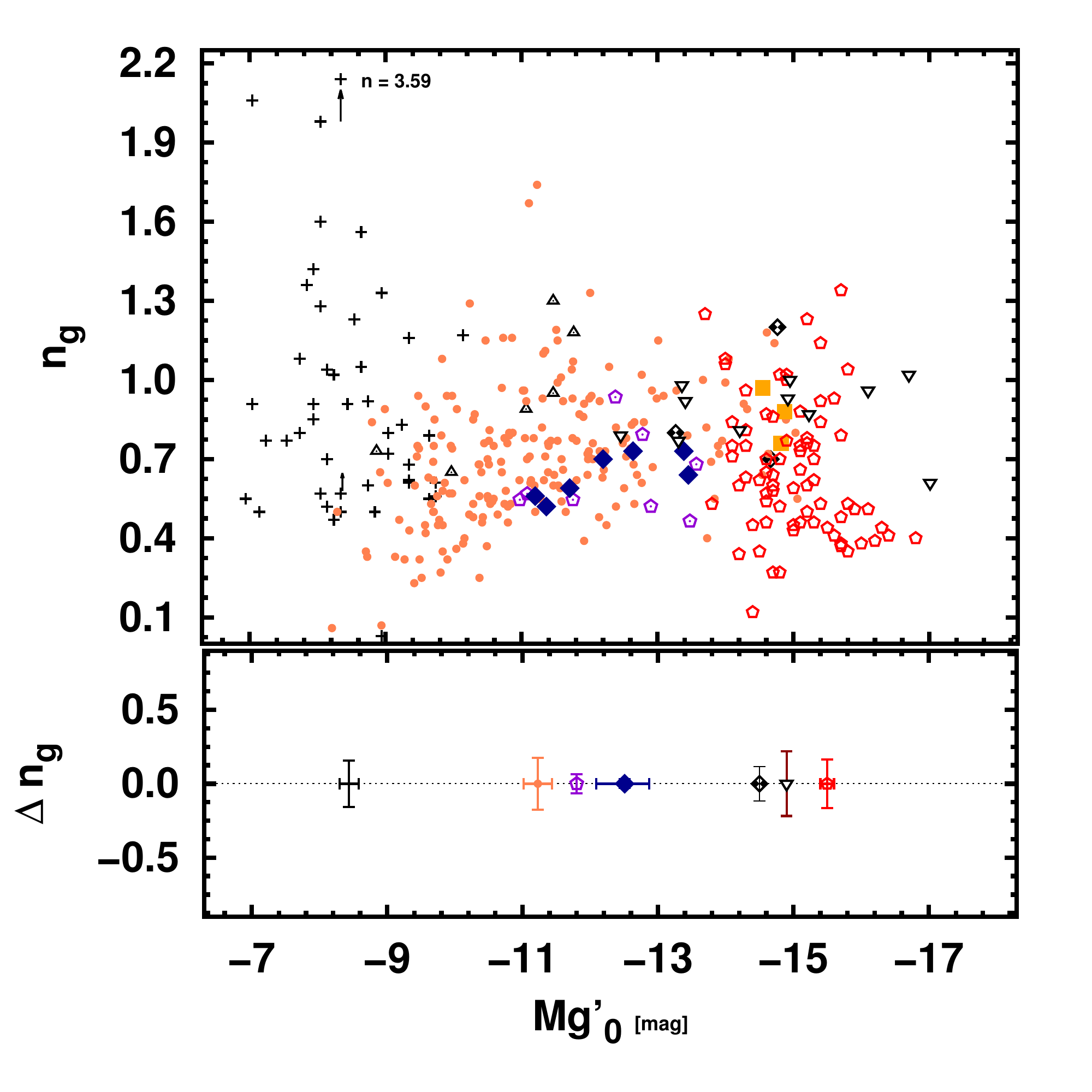}}
\caption{S\'ersic index versus luminosity.  {\it Upper panel:} We show
  the  location  of  different  samples  of LSB  galaxies  and,  as  a
  reference, that  of subsamples of  early-type galaxies of  Virgo and
  NGC\,5044.  LSB galaxies  show a very narrow range of  $n$ values in
  comparison  to those  shown  by typical  early-type galaxies.   {\it
    Lower panel:} Only the LSB galaxy  samples are shown, to allow for
  a more detailed display.  In both panels, symbol code is as given in
  the top margin of the upper panel.}
\label{Index}  
\end{figure}

From Figure\,\ref{Index} we can see that LSB galaxies display S\'ersic
indices  in  a quite  narrow  range  ($n\lesssim2$) in  comparison  to
early-type  galaxies.   The  objects  detected  in  the  direction  of
Pegasus\,I (excepting one galaxy) have brightness profiles that can be
fitted  by a  single  S\'ersic model  with  $0.44 <  n  < 0.97$.   The
remaining   galaxy   presents   two   components:   a   nucleus   with
$(g'-i')=1.01$ mag,  marginally resolved  in the  $g'$, $r'$  and $i'$
images, and an  external component with $1.08 <n<  1.29$, depending on
the filter  (see Table\,\ref{sersic}).  It  is important to  note that
the S\'ersic  index of the  external component is within  the expected
range   for    LSB   galaxies    (e.g   \citealp{2015ApJ...809L..21M};
\citealp{2017A&A...608A.142V}).  Although some  nucleated LSB galaxies
have   been  identified   towards   nearby  groups   such  as   M\,101
\citep{2017ApJ...850..109B} and  Leo\,I \citep{Muller2018}, as  far as
we  know the  role  of nuclei  in connection  with  the formation  and
evolution of LSB galaxies has not been studied in depth.

\section{Discussion and Conclusions} 
\label{discusion}

In this  work we present  eight LSB  galaxies detected in  the central
region  of the  Pegasus\,I group.   Considering their  photometric and
structural properties, these galaxies  show similar characteristics to
those  of   LSB  galaxies   reported  in   the  literature   in  other
environments.  In  particular, we found  that three of  these galaxies
can  be classified  as  UDGs, following  the  criteria established  by
\citet{2015ApJ...798L..45V}:            $\mu_{0,g'}            \gtrsim
24$\,mag\,arcsec$^{-2}$ and  $r_\mathrm{eff}\gtrsim 1.5$\,kpc.   It is
important  to note  that these  criteria do  not imply  that UDGs  are
distinct from the general galaxy population; they simply are large and
extremely diffuse  objects.  Two  of these UDGs  (PEG J232023+081331.4
and    PEG    J232024+081209.0)    are   located    within    Field\,4
(Figure\,\ref{SDSS}), at  a projected distance  of less than  2 arcmin
from  NGC\,7619. The  third one  (PEG J232037+080934.3)  is placed  in
field  7,   at  a   projected  distance  of   $\sim  3$   arcmin  from
NGC\,7626.  The  rest  of  the   detected  LSBs  do  not  present  any
distinctive  peculiarity, excepting  PEG J232054+080838.8  (located in
Field 8  at a projected distance  of $\sim 5$ arcmin  from NGC\,7626),
which displays a nucleus.

Despite  many  efforts  to  understand  the  formation  of  UDGs,  the
observations  have so  far  led to  diverse  scenarios.  For  example,
\citet{2015ApJ...798L..45V}   suggested  that   some  UDGs   could  be
``failed'' galaxies, which  lost their gas after the  formation of the
first generation  of stars, being  strongly dominated by  dark matter.
This scenario  seems to be  in concordance with the  reported velocity
dispersion   of   the   UDG   Dragonfly\,44  in   the   Coma   cluster
\citep{vanDokkum2016ApJ}.  Similarly, the evidence of a high number of
globular clusters  in the UDG Dragonfly  17 supports the idea  that it
could  be a  failed  galaxy with  a  halo mass  similar  to the  Large
Magellanic Cloud \citep{2016ApJ...830...23B}.

  From  a  theoretical  point of  view,  \citet{AmoriscoLoeb2016MNRAS}
  claimed that  UDGs can be  easily explained  by a standard  model of
  disk  formation.  They  suggested that  UDGs are  part of  the dwarf
  galaxy  population   with  a  particularly  high   original  angular
  momentum.  In  this scenario,  the high  angular momentum  makes the
  UDGs more  flat and extended  than typical dEs.   Similarly, through
  Millennium    and   Phoenix    simulations   of    large   clusters,
  \citet{Rong2017}  have shown  that UDGs  are genuine  dwarf galaxies
  that can naturally  emerge from the $\Lambda$CDM  model.  This model
  accurately reproduces the observed properties  of UDGs in the nearby
  clusters. On the other hand, the  existence of UDGs in the field has
  also  been theoretically  predicted  by \citet{DiCintio2017},  using
  cosmological simulations from the  {\it Numerical Investigation of a
    Hundred Astrophysical Objects} (NIHAO)  project.  They showed that
  UDGs  naturally form  in dwarf-sized  halos due  to episodes  of gas
  outflows   associated   with   star   formation.    More   recently,
  \cite{Baushev2018NewA} suggested  that a possible mechanism  for the
  formation  of UDGs  could be  the central  collision of  galaxies in
  their youth stage. This collision would  heat the gas of the system,
  expelling  it from  the galaxies  but without  affecting their  dark
  matter  and  stellar  components,  leaving  them  with  the  typical
  observational properties for this kind of objects.

  Initially, rich clusters  were considered the natural  place to find
  UDGs.   Some of  the studies  that  supported this  idea focused  on
  Fornax                   \citep{Munoz2015ApJ},                  Coma
  \citep{2015ApJ...807L...2K,2015ApJ...798L..45V}       and      Virgo
  \citep{2015ApJ...809L..21M,Davies2016MNRAS}.  More distant groups in
  which     UDGs     have     been    reported     are     Pegasus\,II
  \citep{2017ApJ...846...26S}          and          Abell          168
  \citep{2017MNRAS.468..703R}.  However, there  is recent evidence for
  the existence of  such extended objects outside  rich clusters.  For
  example, additional UDGs have been  detected in HCG\,07, HCG\,25 and
  HCG\,98              \citep{2017MNRAS.468.4039R},             M\,101
  \citep{2016ApJ...833..168M} and HCG\,44 \citep{2016A&A...596A..23S}.
  Moreover, these kind of objects have also been reported in the field
  \citep{2015A&A...581A..82M,2018arXiv180504118G},   underlining   the
  fact that UDGs can be found in diverse environments.

  Following \citet{vanderBurg2016},  UDGs should not survive  near the
  center of clusters, where tidal forces due to the cluster mass would
  prevent their formation. At odds  with this statement, we have found
  three UDG candidates  well inside the central  region of Pegasus\,I,
  very    close   ---in    projection---    to   massive    elliptical
  galaxies.  However,  it  is  worth noticing  the  results  found  in
  Abell\,168 \citep{2017MNRAS.468..703R} and the three isolated groups
  HCG  07/25/98  \citep{2017MNRAS.468.4039R},   where  the  structural
  properties of UDGs change towards the center of the groups showing a
  decrease       in      $r_\mathrm{eff}$,       fainter      $\langle
  \mu_\mathrm{eff}\rangle$, and  larger values  of the  S\'ersic index
  $n$.
  In our case, the most  central UDGs display larger $r_\mathrm{eff}$,
  fainter $\langle \mu_\mathrm{eff}\rangle$, and  lower $n$ values (in
  the $r'$ and  $i'$ filters) than the more distant  one. Anyway, this
  evidence  is  based on  only  three  objects,  and  it is  thus  not
  conclusive.
   In  addition, from  Figure\,\ref{mueff}, it  can be  seen that  LSB
   galaxies reported  by \citet{2017ApJ...846...26S} in  more external
   regions of  Pegasus\,I than those  observed by us,  display similar
   $r_\mathrm{eff}$ than those  of the central regions,  but they show
   brighter  $\langle \mu_\mathrm{eff}\rangle$  (these authors  do not
   provide $n$ values  for these galaxies).  Considering  all the UDGs
   in Pegasus\,I reported so far, it  is found that they spread over a
   wide range of colors both in $(g'-i')$ and $(g'-r')$, covering both
   the red and the blue sequences.  No correlation is found among UDGs
   between   color    and   the   environment   where    they   reside
   \citep{2017MNRAS.468..703R,2017ApJ...846...26S}.
   
  In  general, LSB  galaxies show  smooth surface  brightness profiles
  that are well characterized  by single-component S\'ersic functions,
  with index $n  \lesssim 2$. In particular, the  galaxies detected in
  the  direction of  Pegasus\,I  display a  more  restricted range  of
  values ($0.44 <  n < 0.97$).  It is interesting  to note that, while
  typical  early-type galaxies  follow a  well defined  luminosity-$n$
  relation  \citep{1993MNRAS.265.1013C,1994ApJS...93..397C},  the  LSB
  galaxies,  in  contrast, display  $n$  values  scattering around  $n
  \lesssim 1$, showing no trend with luminosity.  There is an opposite
  situation   if    we   consider   the   color--magnitude    or   the
  $\langle{\mu}_{{\rm eff}}\rangle$--luminosity  planes, where typical
  early-type  galaxies display  well  delineated  sequences while  LSB
  galaxies  show  no clear  trend  at  all.  In  particular,  high-$n$
  profiles  are  usually  explained   by  major  mergers  and  violent
  relaxation or by numerous minor  mergers, while low-$n$ profiles are
  usually  associated  to  secular evolution  \citep[][and  references
    therein]{2008ApJ...688...67E}.   In this  context,  and given  the
  remarkably different  behavior between  the $n$ values  displayed by
  typical early-type  galaxies and  LSB galaxies (regardless  of their
  sizes), it  can be proposed  that both  types of systems  arise from
  different   formation  processes,   which  also   implies  different
  formation time  scales (rapid and slow,  respectively). Two distinct
  formation  pathways  might  also  explain  the  different  behaviors
  observed  between  other (color-chemical,  $r_{\rm  eff}$-dynamical)
  physical parameters. However, we cannot  rule out that some of these
  objects   may  have   a  tidal   origin.   In   the  cases   of  PEG
  J232023+081331.4, PEG J232024+081209.0  and PEG J232049+080806.3, we
  are able to identify some extended and very diffuse substructures in
  their images  that might be  associated with this kind  of processes
  (see  Figure\,\ref{J232023+081331.4}, Figure\,\ref{J232024+081209.0}
  and Figure\,\ref{J232049+080806}).  One of these structures presents
  $\mu_{r'}\gtrsim 27~$mag~arcsec$^{-2}$.
     
  Another point to note is the correlation that seems to exist between
  the effective radii of LSB galaxies  and the distances of their host
  clusters/groups,  in the  sense that  the  actual sizes  of the  LSB
  galaxies increase with  distance.  This correlation could  only be a
  selection effect due  to the limitations we have to  detect such low
  surface brightness  objects at  large distances.  Small  distant LSB
  galaxies can be confused with background objects, while large nearby
  LSB  galaxies might  not be  detected because  of their  exceedingly
  large angular sizes  (although the probability of  finding large LSB
  galaxies in the a small volume  around the Local Group must be quite
  low).  Recently,  \citet{Muller2018} found possible  extremely large
  UDGs in  the Leo\,I group ($D  \sim 10.7$\,Mpc).  If these  UDGs are
  confirmed to be Leo\,I members, they would be some of the closest.

  Within this context, where  the formation scenarios and evolutionary
  paths followed by these objects  remain speculative, the database of
  observed LSB galaxies in  different environments should be enlarged,
  thus  providing  the  basis  to  better  constrain  models  of  LSBs
  formation and evolution.
  As a  future work,  we expect  to obtain  spectroscopic data  of LSB
  galaxies  to  confirm their  membership  to  Pegasus\,I.  This  will
  certainly help us to unveil the real nature of these extreme stellar
  systems.

\begin{acknowledgements}

Based  on observations  obtained  at the  Gemini Observatory  (Program
GN-2008B-Q-14,   GN-2012A-Q-55,   GN-2014A-Q-70,   GN-2014B-Q-17   and
GN-2015B-Q-13), which  is operated by the  Association of Universities
for Research  in Astronomy, Inc.,  under a cooperative  agreement with
the  NSF on  behalf of  the Gemini  partnership: the  National Science
Foundation (United  States), the  National Research  Council (Canada),
CONICYT   (Chile),  Minist\'{e}rio   da   Ci\^{e}ncia,  Tecnologia   e
Inova\c{c}\~{a}o (Brazil) and Ministerio  de Ciencia, Tecnolog\'{i}a e
Innovaci\'{o}n Productiva (Argentina).\\

This research  has made  use of  the NASA/IPAC  Extragalactic Database
(NED) which is  operated by the Jet  Propulsion Laboratory, California
Institute of Technology, under  contract with the National Aeronautics
and Space Administration.\\

Funding  for  SDSS-III  has  been  provided by  the  Alfred  P.  Sloan
Foundation,  the  Participating  Institutions,  the  National  Science
Foundation, and the U.S. Department  of Energy Office of Science.  The
SDSS-III web site is http://www.sdss3.org/.\\

SDSS-III is managed  by the Astrophysical Research  Consortium for the
Participating Institutions of the SDSS-III Collaboration including the
University of  Arizona, the Brazilian Participation  Group, Brookhaven
National  Laboratory,   Carnegie  Mellon  University,   University  of
Florida,  the French  Participation  Group,  the German  Participation
Group, Harvard  University, the Instituto de  Astrofisica de Canarias,
the Michigan State/Notre Dame/JINA  Participation Group, Johns Hopkins
University,   Lawrence  Berkeley   National  Laboratory,   Max  Planck
Institute for Astrophysics, Max  Planck Institute for Extraterrestrial
Physics, New Mexico State University,  New York University, Ohio State
University, Pennsylvania  State University, University  of Portsmouth,
Princeton University,  the Spanish Participation Group,  University of
Tokyo,  University  of  Utah,  Vanderbilt  University,  University  of
Virginia, University of Washington, and Yale University.  \\

This  work   was  funded   with  grants   from  Consejo   Nacional  de
Investigaciones Cient\'{\i}ficas y T\'ecnicas and Universidad Nacional
de La Plata (Argentina).

\end{acknowledgements}

%
%
\newpage
\clearpage

\bibliographystyle{aa}
\bibliography{manuscrito_Gonzalezetal_revised}
  
\appendix
\section{LSB galaxies identified in Pegasus\,I}
\label{apendice}
In this section we present similar figures to Figure\,\ref{J231956+081253},
for the rest of the extended LSB galaxies
identified      in      Pegasus\,I.
  
%
\begin{figure}[h!]
\center
\hspace{0.55cm}
\includegraphics[scale=0.25]{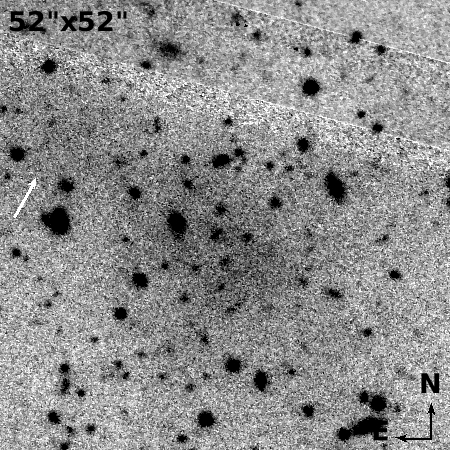}
\includegraphics[scale=0.25]{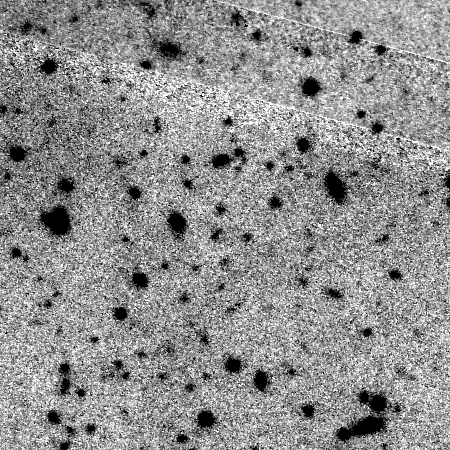}
\resizebox{1.0\hsize}{!}{\includegraphics{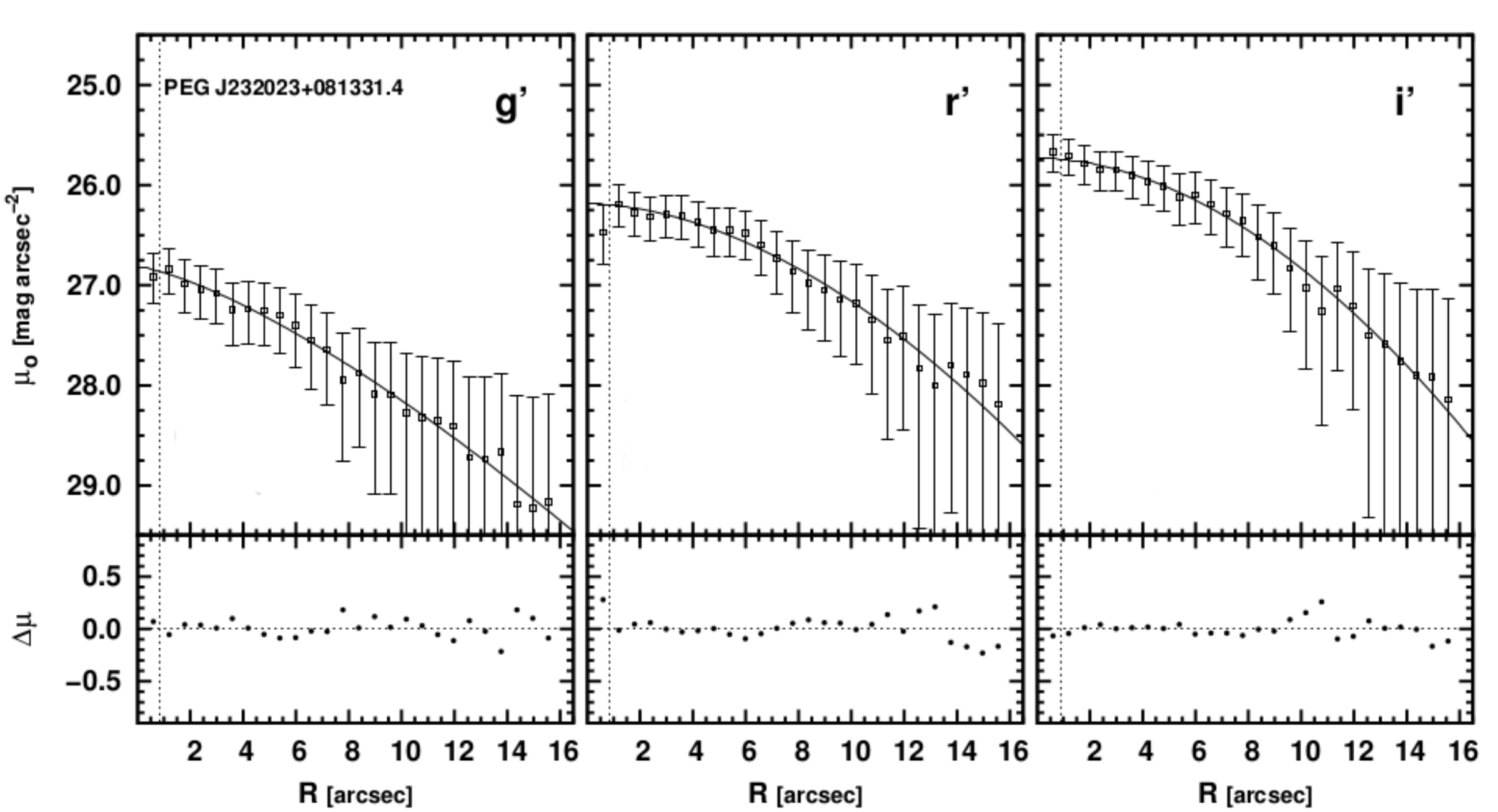}}\\
\caption{Idem    Figure\,\ref{J231956+081253}     but    for    galaxy
  PEG\,J232023+081331.4. In this  case, the size of the  frames in the
  upper  panels is  52  $\times$  52 arcsec.   There  seems  to be  an
  extremely faint,  low-surface-brightness structure connected  to the
  galaxy {\it (white  arrow)}.  This structure might be  evidence of a
  tidal  origin for  the galaxy,  or of  an interaction  with a  tidal
  structure.   This  tidal structure  may  be  affecting the  galaxy's
  colors.  The  galaxy is superposed  on the dithering pattern  of the
  gaps between the CCDs, resulting in larger photometric errors.}
\label{J232023+081331.4}
\end{figure}

\begin{figure}[h!]
\center
\hspace{0.55cm}
\includegraphics[scale=0.245]{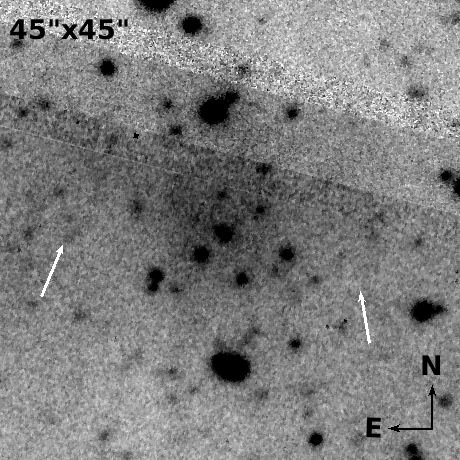}
\includegraphics[scale=0.245]{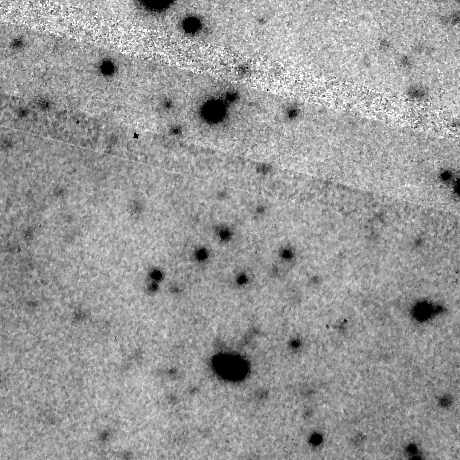}
\resizebox{1.0\hsize}{!}{\includegraphics{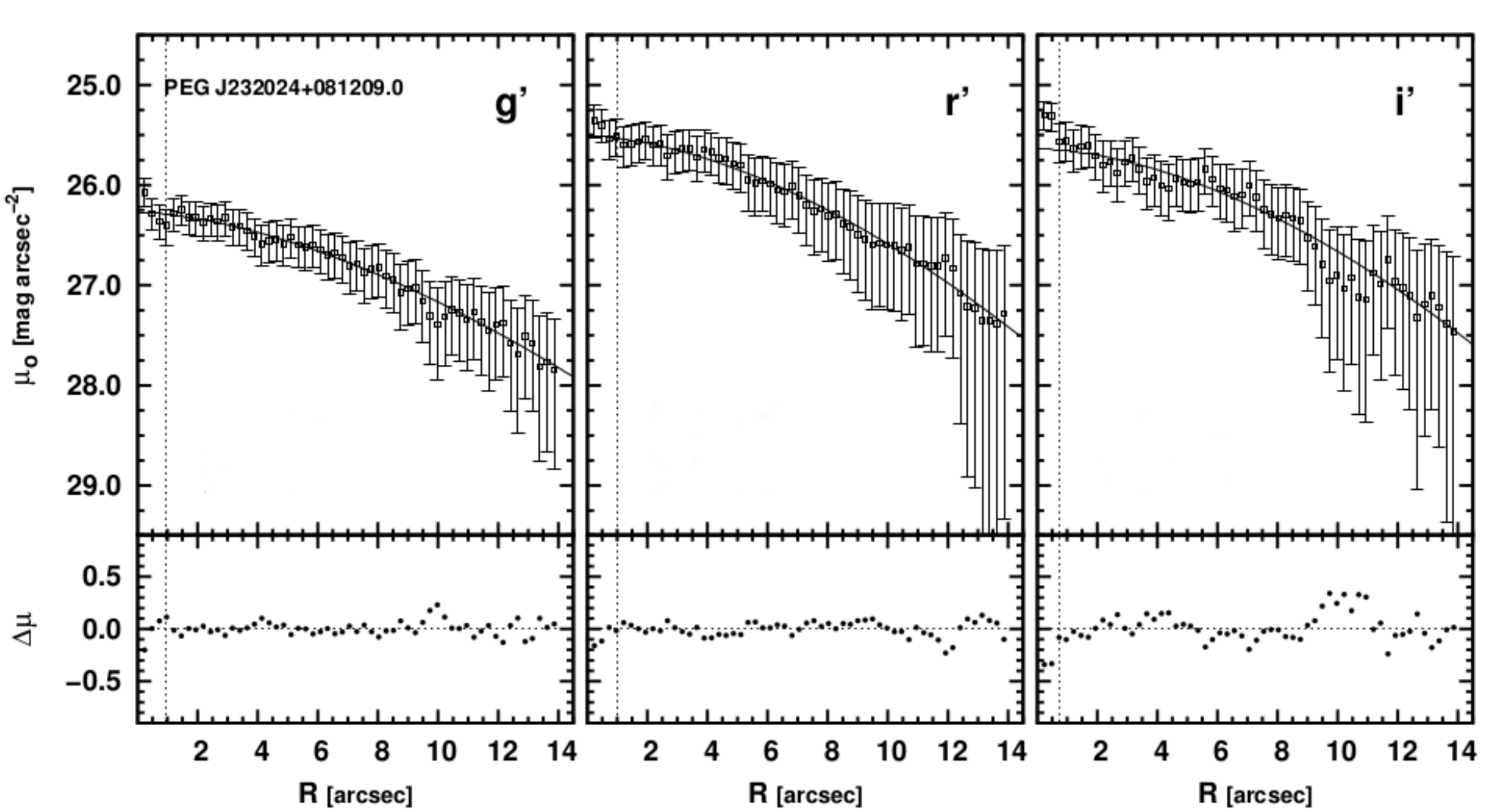}}\\
\caption{Idem   Figure\,\ref{J231956+081253}   but  for   galaxy   PEG
  J232024+081209.0. In this case, the size  of the frames in the upper
  panels   is   45   $\times$   45  arcsec.    an   extremely   faint,
  low-surface-brightness structure connected to the galaxy {\it (white
    arrows)}.  This structure might be  evidence of a tidal origin for
  the galaxy, or of an interaction with a tidal structure.  This tidal
  structure  may be  affecting  the galaxy's  colors.   The galaxy  is
  superposed on  the dithering pattern  of the gaps between  the CCDs,
  resulting in relatively large photometric errors.}
\label{J232024+081209.0}
\end{figure}

\begin{figure}[h!]
\center
\hspace{0.55cm}
\includegraphics[scale=0.25]{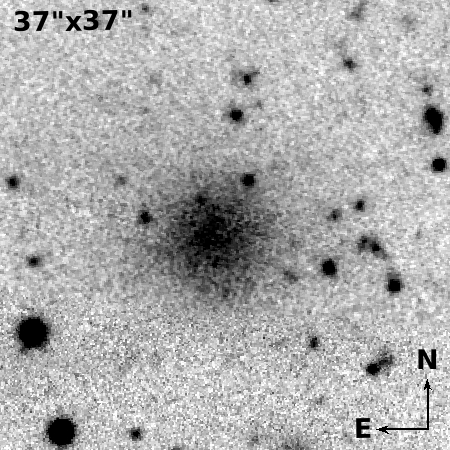}
\includegraphics[scale=0.25]{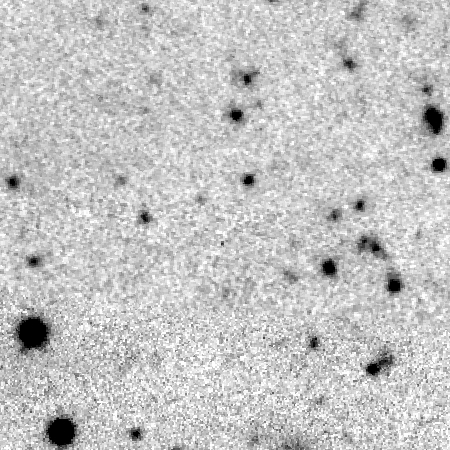}
\resizebox{1.0\hsize}{!}{\includegraphics{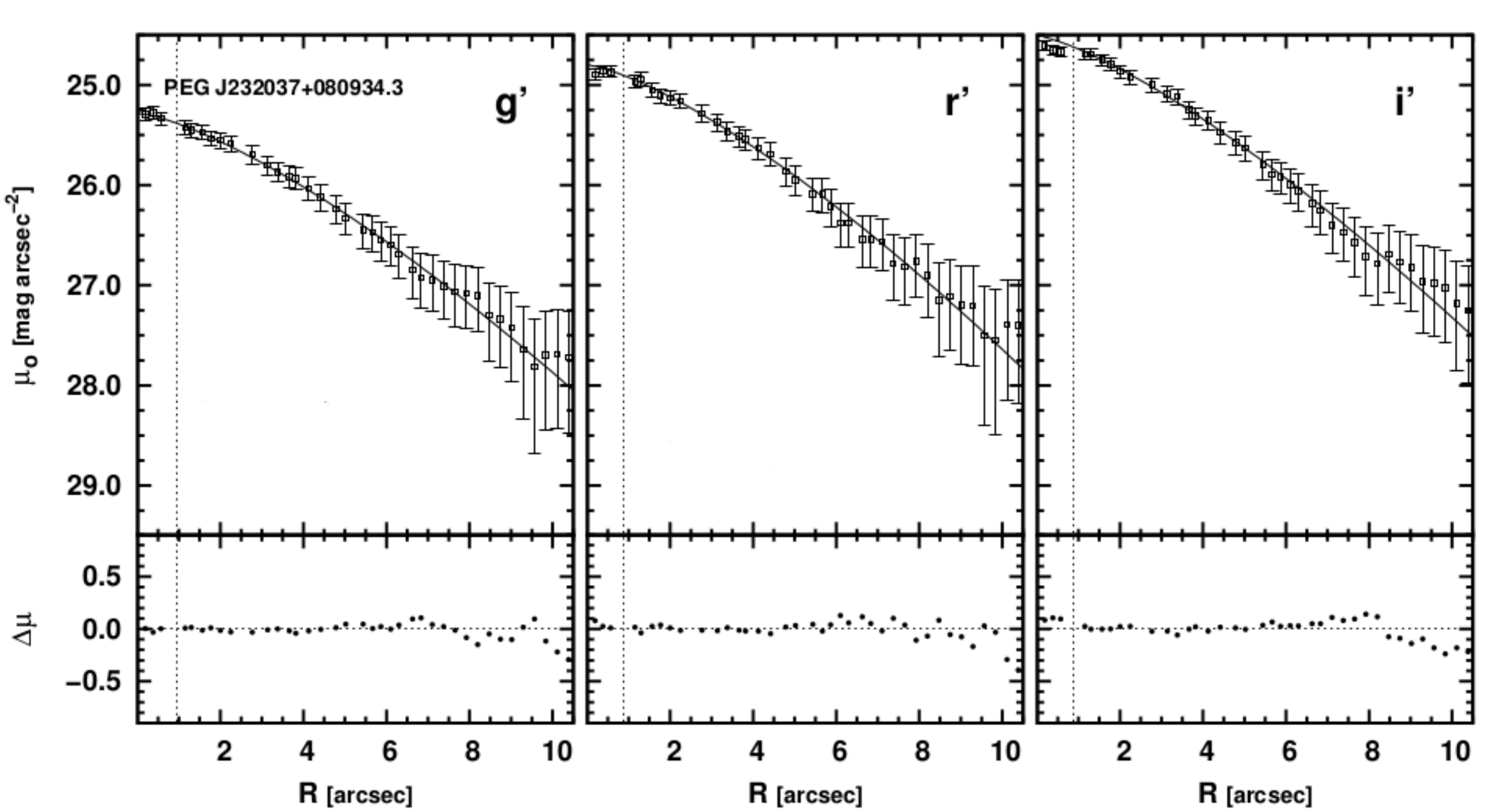}}\\
\caption{Idem    Figure\,\ref{J231956+081253}     but    for    galaxy
  PEG\,J232037+080934.3. In this  case, the size of the  frames in the
  upper panels is 37 $\times$ 37 arcsec.}
\label{J232037+080934.3}
\end{figure}

\begin{figure}[h!]
\center
\hspace{0.55cm}
\includegraphics[scale=0.25]{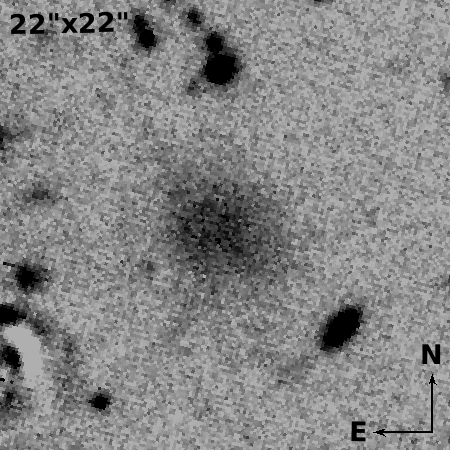}
\includegraphics[scale=0.25]{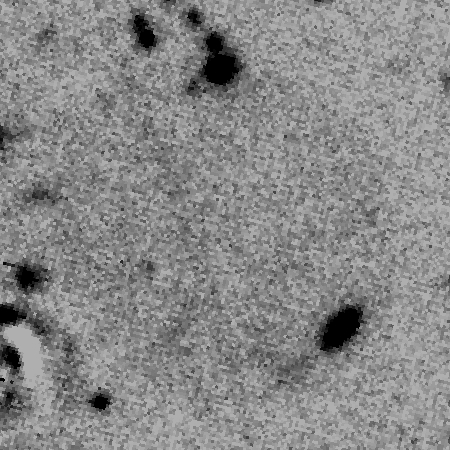}
\resizebox{1.0\hsize}{!}{\includegraphics{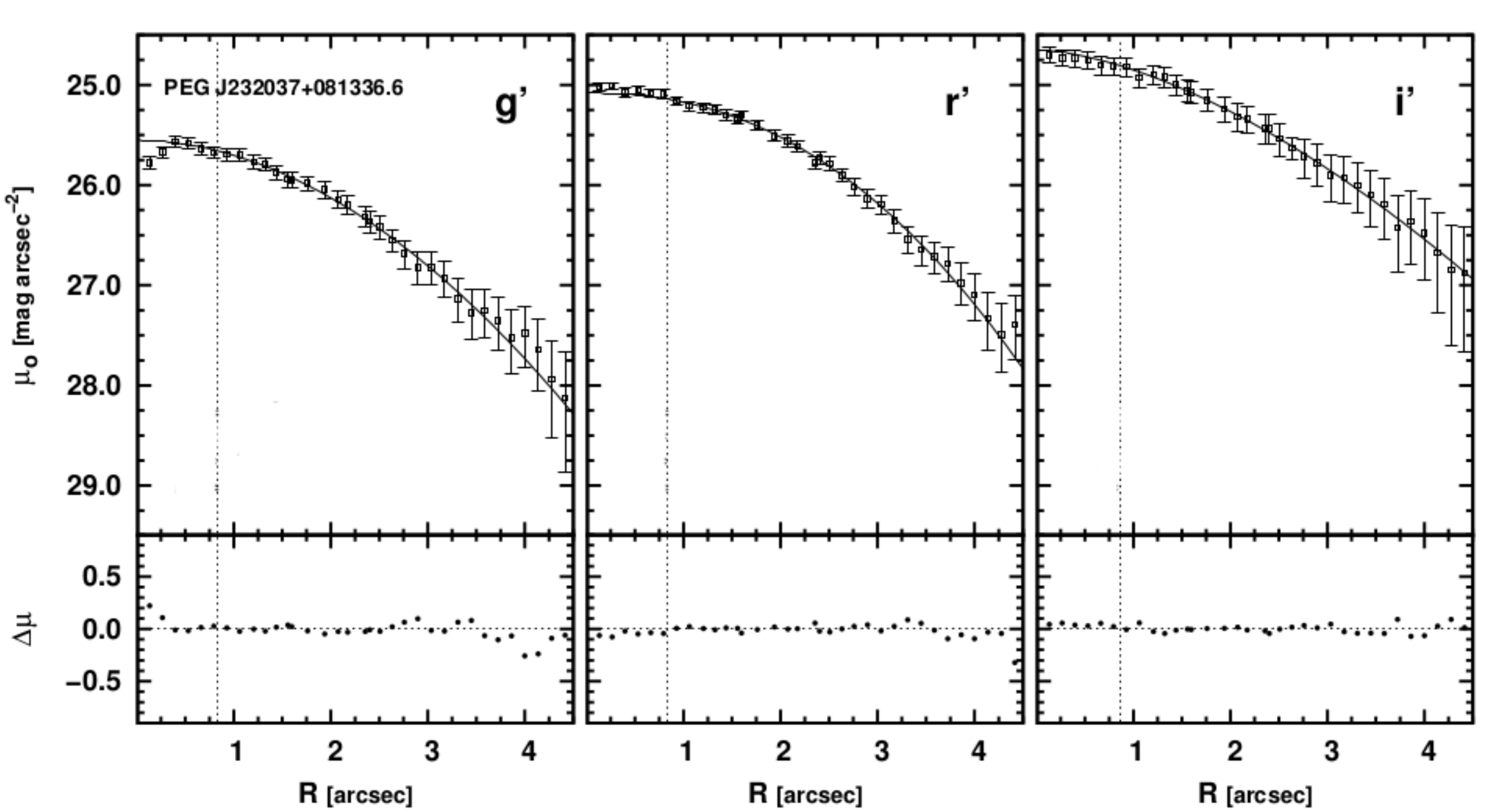}}\\
\caption{Idem    Figure\,\ref{J231956+081253}     but    for    galaxy
  PEG\,J232037+081336.6. In this  case, the size of the  frames in the
  upper panels is 22 $\times$ 22 arcsec.}
\label{J232037+081336.6}
\end{figure}

\begin{figure}[h!]
\center
\hspace{0.55cm}
\includegraphics[scale=0.25]{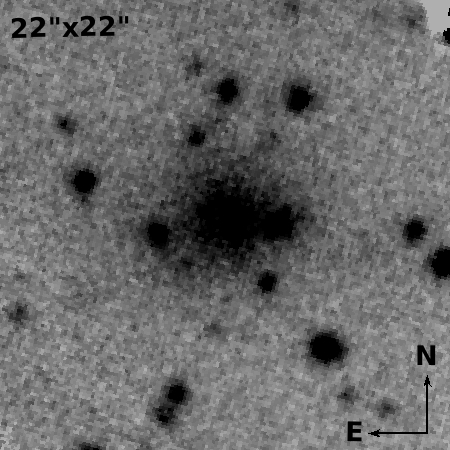} 
\includegraphics[scale=0.25]{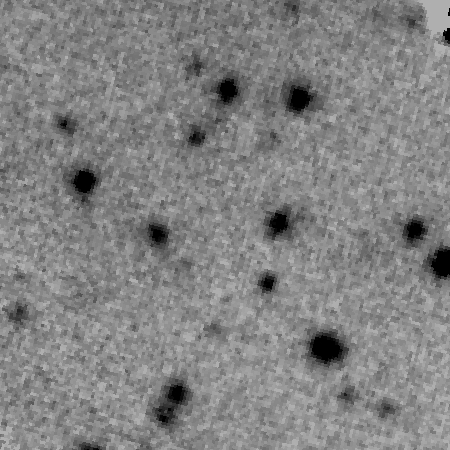}
\resizebox{1.0\hsize}{!}{\includegraphics{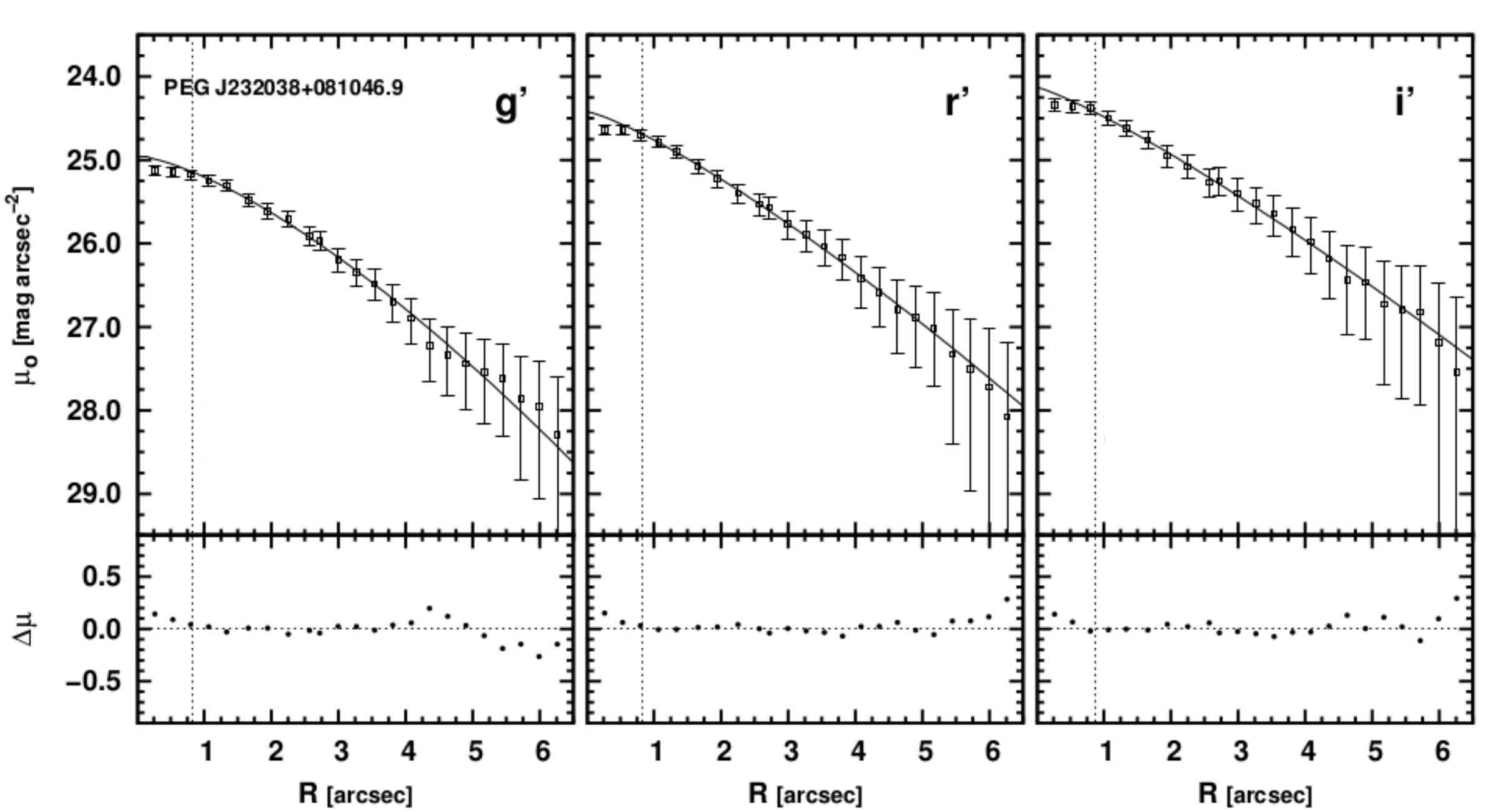}}\\
\caption{Idem    Figure\,\ref{J231956+081253}     but    for    galaxy
  PEG\,J232038+081046.9. In this  case, the size of the  frames in the
  upper panels is 22 $\times$ 22 arcsec.}
\label{J232038+081046.9}
\end{figure}

\begin{figure}[h!]
\center
\hspace{0.55cm}
\includegraphics[scale=0.265]{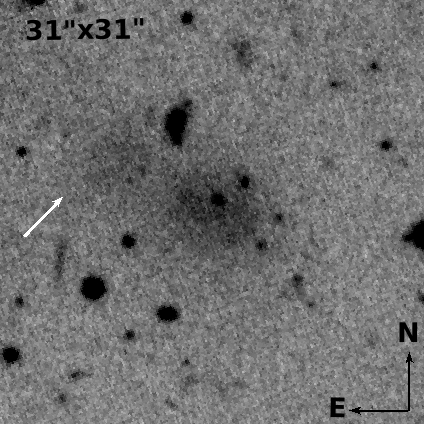}
\includegraphics[scale=0.265]{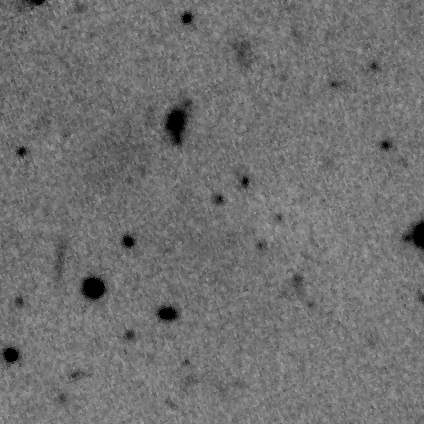}
\resizebox{1.0\hsize}{!}{\includegraphics{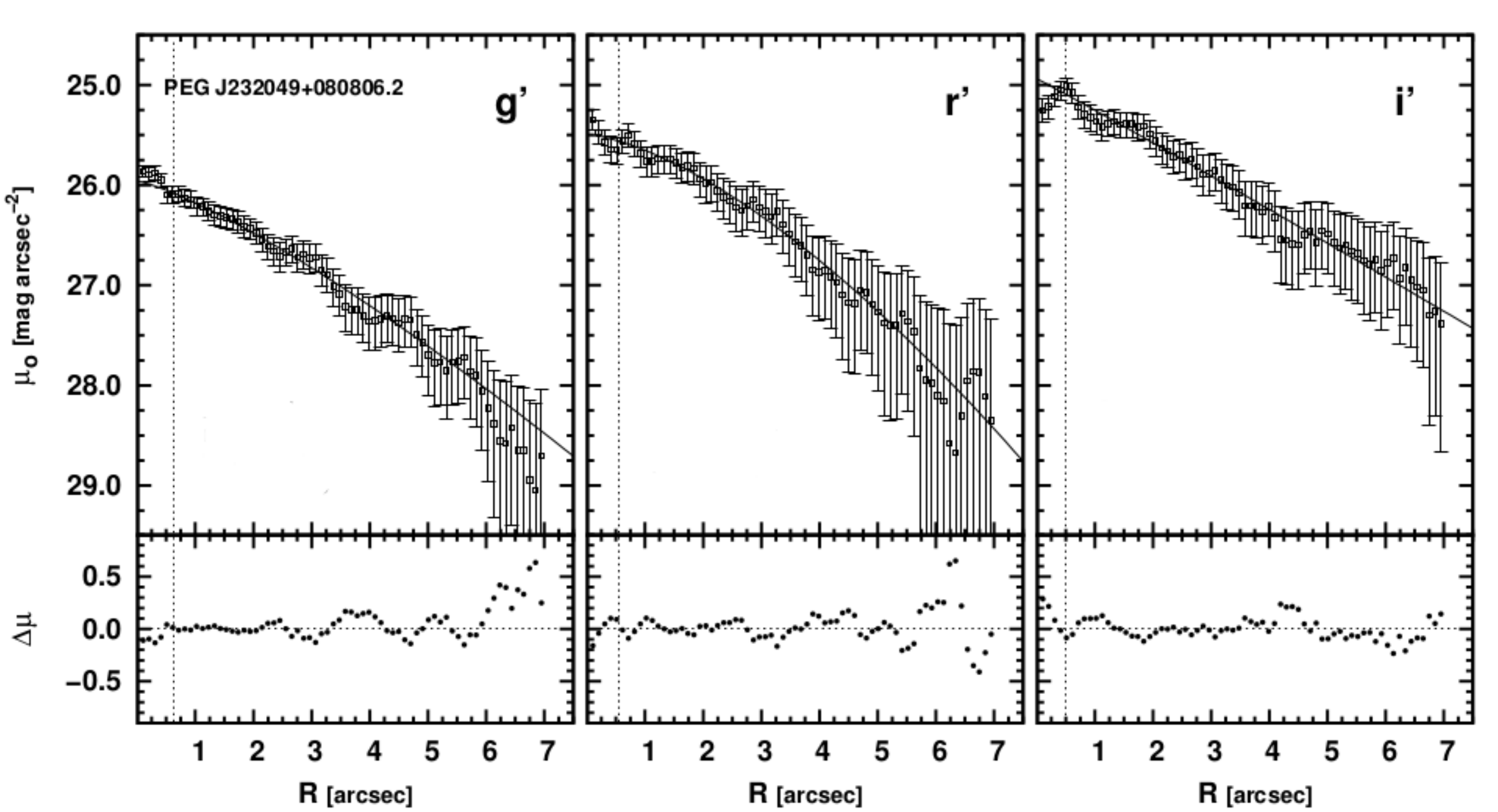}}\\
\caption{Idem    Figure\,\ref{J231956+081253}     but    for    galaxy
  PEG\,J232049+080806.3. In this  case, the size of the  frames in the
  upper  panels is  31  $\times$  31 arcsec.   There  seems  to be  an
  extremely   faint   low-surface-brightness   object   or   structure
  ($\mu_{r'} \gtrsim  27~$mag~arcsec$^{-2}$) to  the northwest  of the
  galaxy  {\it  (white arrow)}.  Quite  speculatively,  this might  be
  evidence of a tidal origin for this object .}
\label{J232049+080806}
\end{figure}

\begin{figure}[h!]
\center
\hspace{0.55cm}
\includegraphics[scale=0.25]{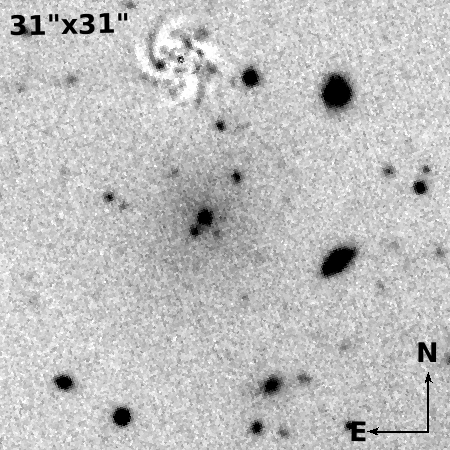}
\includegraphics[scale=0.25]{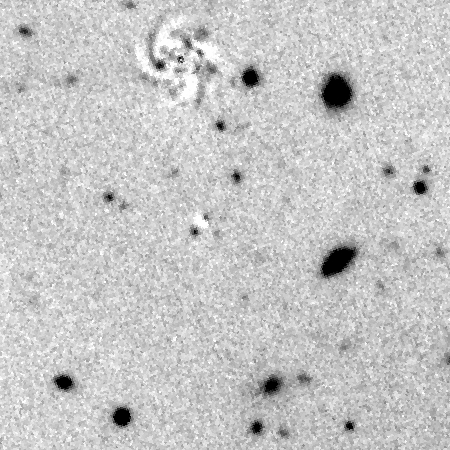}
\resizebox{1.0\hsize}{!}{\includegraphics{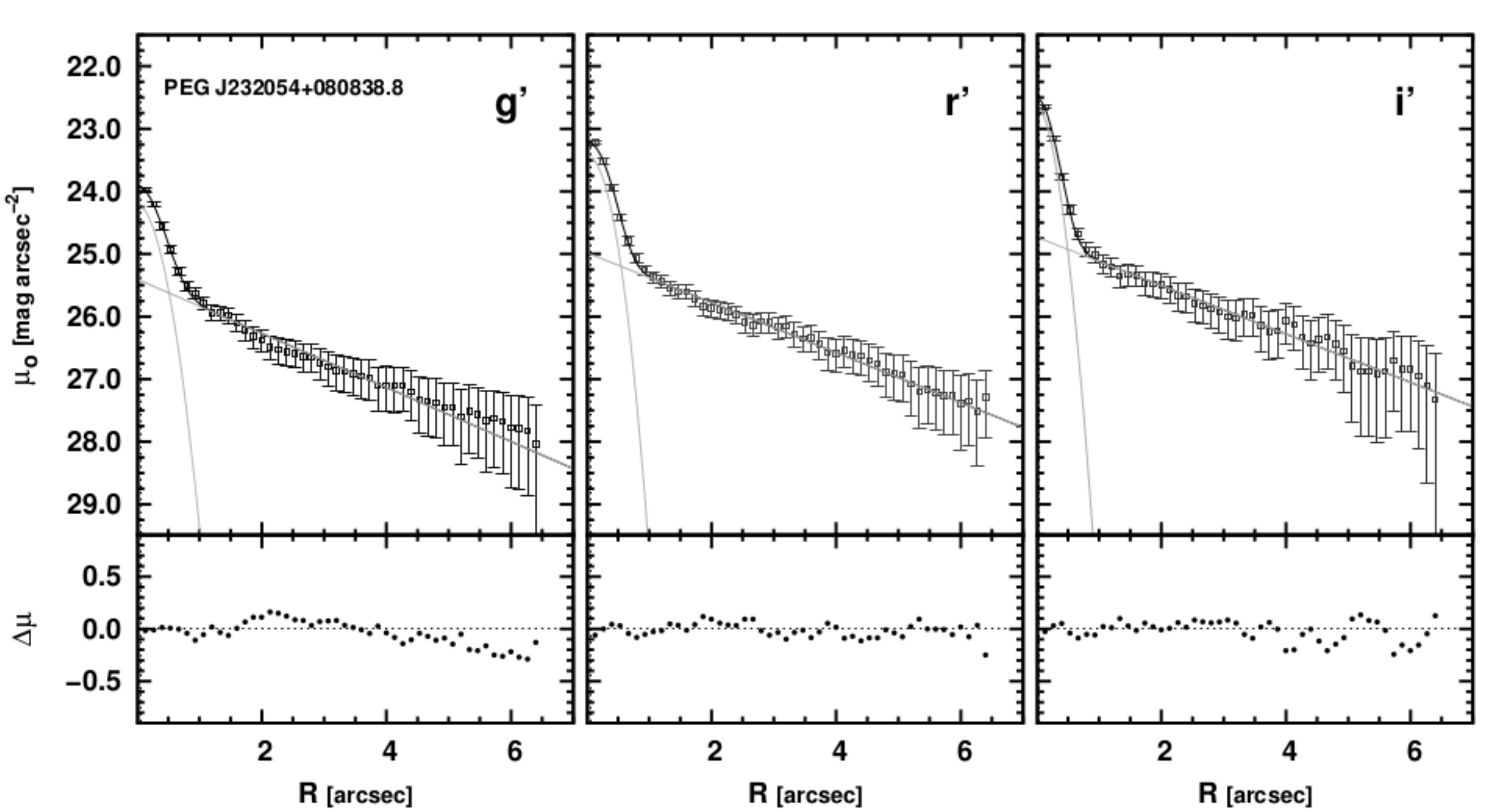}}\\
\caption{Idem    Figure\,\ref{J231956+081253}     but    for    galaxy
  PEG\,J232054+080838.8. In this  case, the size of the  frames in the
  upper  panels is  31 $\times$  31  arcsec.  This  object presents  a
  nucleus which is marginally resolved in the $g'$, $r'$ and $i'$-band
  images.}
\label{J232054+080838}
\end{figure}

\end{document}